%% file: dampe.tex
\documentclass[aps,prl,twocolumn,a4paper]{revtex4}

\usepackage{amsmath,amssymb} 
\usepackage{bm} %
 
\usepackage{color} %
\usepackage{graphicx}

\usepackage[colorlinks]{hyperref} %
\hypersetup{
citecolor=blue,
linkcolor=blue,
urlcolor=blue
}

\usepackage{epstopdf}
\usepackage{multirow}

\usepackage{comment}

\newcommand{\note}[1]{#1}{\ignorespacesafterend}

\newcommand{\fig}[1]{Fig.~\ref{#1}}

\newcommand{\tab}[1]{Tab.~\ref{#1}}

\newcommand{\eq}[1]{Eq.~(\ref{#1})}

\newcommand{\tx}[1]{\text{#1}}
	
\newcommand{\sigmav}{\langle \sigma v \rangle}

\begin{document}
\title{Origins of  sharp cosmic-ray electron  structures and the DAMPE excess}
\author{Xian-Jun Huang$^{a,b}$}
\author{Yue-Liang Wu$^{a}$}
\author{Wei-Hong Zhang$^{a}$}
\author{Yu-Feng Zhou$^{a}$}
\affiliation{$^{a}$
	CAS Key Laboratory of Theoretical Physics, 
	Institute of Theoretical Physics, Chinese Academy of Sciences,
	Beijing, 100190, China,
	}
\affiliation{
	University of Chinese Academy of Sciences, 
	Beijing 100049, China,
}
\affiliation{$^{b}$
 Sichuan University of Science and Engineering, Zigong 643000, China.}
\begin{abstract}
Nearby sources  may contribute to cosmic-ray electron  (CRE) structures 
at high energies.
Recently, the first  DAMPE results on the CRE flux hinted at 
a  narrow  excess 
at   energy $\sim 1.4$~TeV.  
We show that in general a  spectral structure with a narrow width 
appears  in two scenarios:
I) ``{\it Spectrum broadening}'' for the continuous sources 
with a $\delta$-function-like  injection spectrum.
In this scenario, %
a finite width can develop after propagation through the Galaxy,
which
can reveal %
the distance of the source.
Well-motivated   sources include mini-spikes and subhalos formed 
by dark matter (DM) particles $\chi_{s}$ which annihilate directly into $e^{+}e^{-}$ pairs.
II) ``{\it Phase-space shrinking}'' for burst-like sources with a power-law-like injection spectrum.
The  spectrum after propagation can shrink at a cooling-related cutoff energy 
and form a sharp spectral peak.
The peak can be  more prominent due to the energy-dependent diffusion.
In this scenario, the width of the excess constrains both 
the power index and the distance of the source.
Possible such sources are pulsar wind nebulae (PWNe) and supernova remnants (SNRs).
We analysis the DAMPE excess and find that 
the continuous DM sources should be fairly close within $\sim 0.3$~kpc, 
and the annihilation cross sections are close to 
the thermal value.
For the burst-like source, 
the narrow width of the excess suggests that 
the injection spectrum must be  hard 
with power index significantly less than two, 
the distance is within $\sim(3-4)$~kpc, and
the age of the source is  $\sim 0.16$~Myr.
In both scenarios,  
large  anisotropies  in the CRE flux are predicted. 
We identify possible candidates of mini-spike and PWN sources
in the current Fermi-LAT 3FGL and ATNF catalog, respectively.
The diffuse $\gamma$-rays from these sources
can be well below the Galactic diffuse $\gamma$-ray backgrounds
and less constrained by the Ferm-LAT data,
if they are located at the low Galactic latitude regions.	
\end{abstract}
\pacs{xx.xx}
\preprint{ [\today ]}
\maketitle %

\paragraph{\bf Introduction.}
Cosmic-ray (CR)  electrons and positrons (CREs)
with energies above TeV  plays an important role in understanding 
the nearby origins of CRs within a few kpc
\cite{Shen:1970apj}.
Structures in the energy spectrum of CREs are expected, 
if the nearby sources are dominated by one or a few discrete sources.
The current space experiments have begun to directly probe  this energy region.
For instance, 
the AMS-02~\cite{Aguilar:2014fea}, 
Fermi-LAT~\cite{Abdollahi:2017nat} and 
CALET~\cite{Adriani:2017efm} experiments have measured 
the  flux of CRE up to 1, 2 and 3~TeV, respectively,  
without observing any significant structures so far.
Recently, 
the DAMPE experiment has reported  the first high energy resolution 
measurement of  the CRE flux up to  4.6~TeV
\cite{Chang:2017xx}.
The measured energy spectrum of CRE 
steepened above $\sim0.9$~TeV,
consistent with the  results from 
the ground-based atmospheric Cherenkov telescopes
\cite{
	Aharonian:2008aa,%
	Aharonian:2009ah,%
	HESS:2017ICRC,%
	BorlaTridon:2011dk,%
	Staszak:2015kza%
}.
Of interest, the DAMPE data  also hinted at an excess over  
the expected background  in a narrow energy interval  $\sim(1.3-1.5)$~TeV.
Making use of the  DAMPE data in the energy range 55~GeV--4.6~TeV, and 
assuming  a broken power-law background flux,
we find that the local and global significance of the possible narrow excess is
$\sim 3.7~\sigma$ and $\sim2.5~\sigma$, respectively
(details of the data analysis are shown in the supplementary material).
%

\note{
In light of the possible DAMPE ``excess'',
it is of general interest to address the question of 
what kind of sources are responsible for a sharp
spectral feature in CRE flux.}
In this work, 
we explore the origins of a sharp spectral structure and 
emphasize  that  the space-time location of  source can be inferred from 
the spectral feature of the CRE flux.
We show that in general a sharp spectral structure can be produced   in two  complementary scenarios:
I)
for  continuous sources with a line-shape injection spectrum, 
a  finite width can develop after propagation in the Galaxy
(dubbed ``spectrum broadening'').
Well-motivated  sources are nearby DM substructures 
such as mini-spikes and DM subhalos of DM particles $\chi_{s}$
with  $e^{+}e^{-}$ the dominant annihilation final states.
In this scenario, the spectral shape or the width of the excess can be used to estimate  
the distance to the source.
II)
for burst-like sources with a power-law injection spectrum,
the spectrum after propagation can shrink at a cooling-related cutoff energy 
and form a sharp spectral peak  (dubbed ``phase-space shrinking'').
Energy-dependent diffusion also contributes  to the spectral rising. 
Typical  sources of this type are pulsar wind nebulae (PWNe) and 
supernova remnants (SNRs).
In this scenario, the power index and the distance of the source are strongly 
constrained by the width of the excess.

In view of the DAMPE excess, we find: 
\romannumeral 1)
for the continuous sources, 
the favoured distance should be less than $\sim 0.3$~kpc.
The source can be  ``mini-spikes'' or DM subhalos  with 
the favoured DM annihilation cross section around the typical thermal value.
\romannumeral 2)
For the burst-like source, 
the injection spectrum must be hard with power index significantly below two, and
the distance within $\sim(3-4)$~kpc. 
The age of the source is determined to be $\sim0.16$~Myr.
\romannumeral 3) 
For both sources,
large anisotropies in the arrival direction of the CRE are predicted,
which are close to the current Ferm-LAT upper limits.
\romannumeral 4) 
We identify possible candidates for  mini-spikes (PWNe)
from the catalogue of Fermi-LAT  3GFL 
(ATNF pulsar catalogue).
\begin{figure}[thb]
	\centering
	\includegraphics[width=0.48\columnwidth]{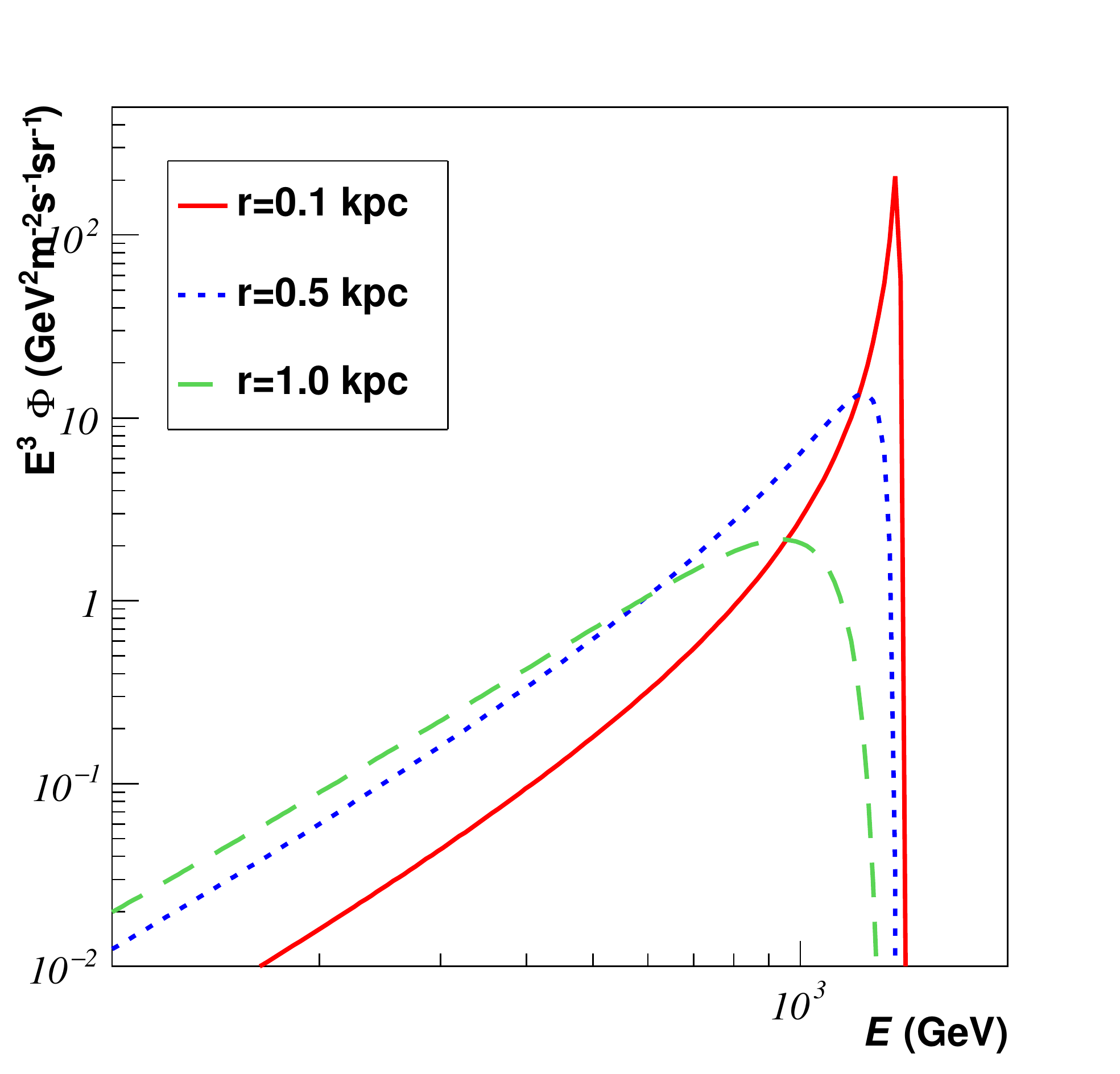}
	\includegraphics[width=0.48\columnwidth]{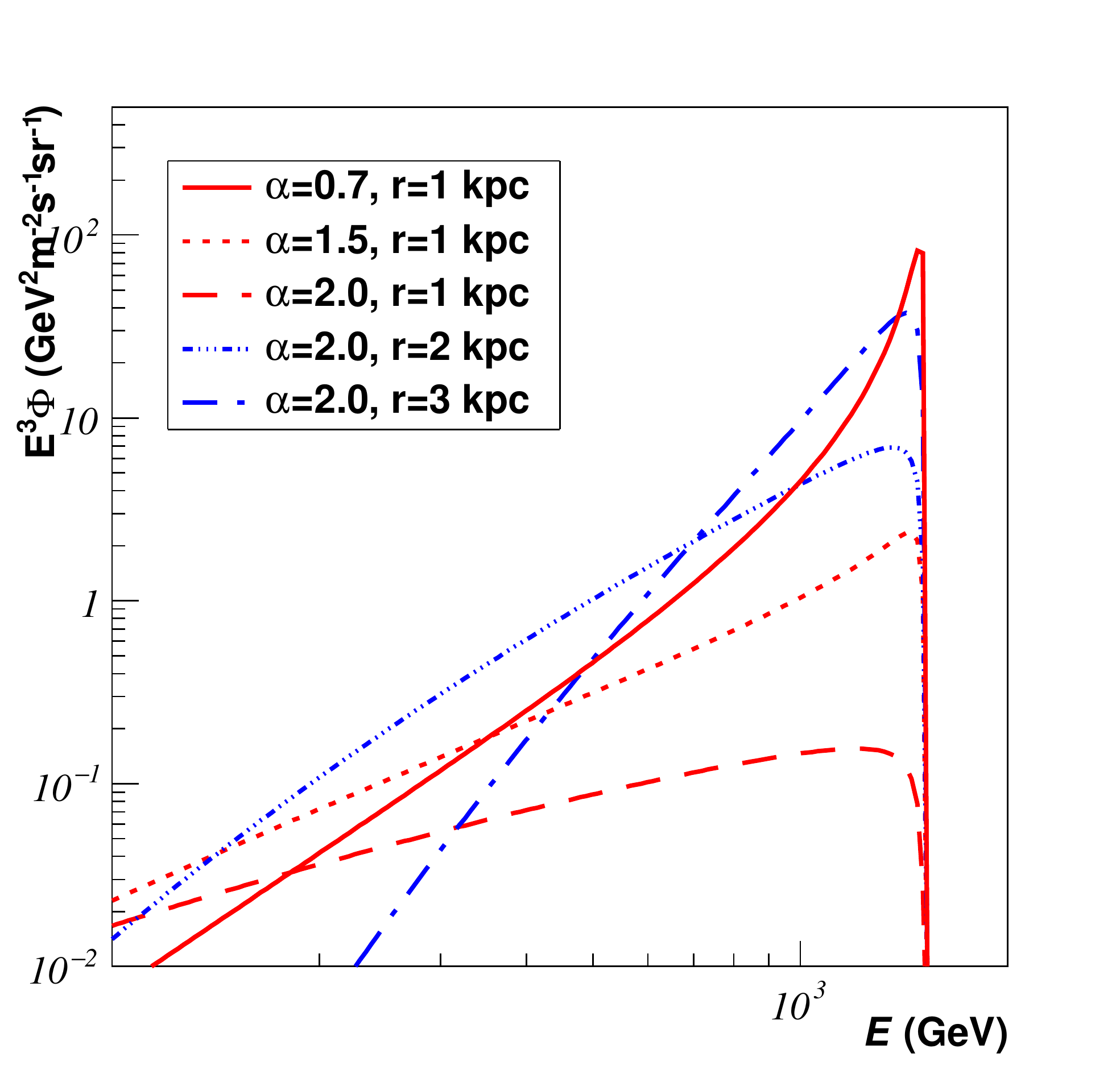}
	\caption{		
		Left) 
		Effect of  ``spectrum broadening'' on CRE flux from 
		a continuous point-like source described in 
		\eq{eq:continuous-point-solution} with a growing distance $r=0.1-1.0$~kpc.		 
		$E_{0}$ is fixed at 1.4~TeV.
		Right)
		Effect of ``phase-space shrinking'' for the burst-like sources in 
		\eq{eq:burst--solution} with  a decreasing $\alpha$ from 2.0 to 0.7 and fixed $r=1$~kpc,
		and ``energy-dependent diffusion''  for  a growing  distance $r=1-3$~kpc with fixed $\alpha=2$.
		The age of the source is  fixed at $t=0.15$~Myr.
		For both sources, the spectra are normalized to a total flux
		$\Phi=10^{-9}\mbox{m}^{-2}\mbox{s}^{-1}\mbox{sr}^{-1}$.
	}\label{fig:spectral-feature}
\end{figure}
\begin{figure*}[htb]
	\centering
	\includegraphics[width=0.95\textwidth]{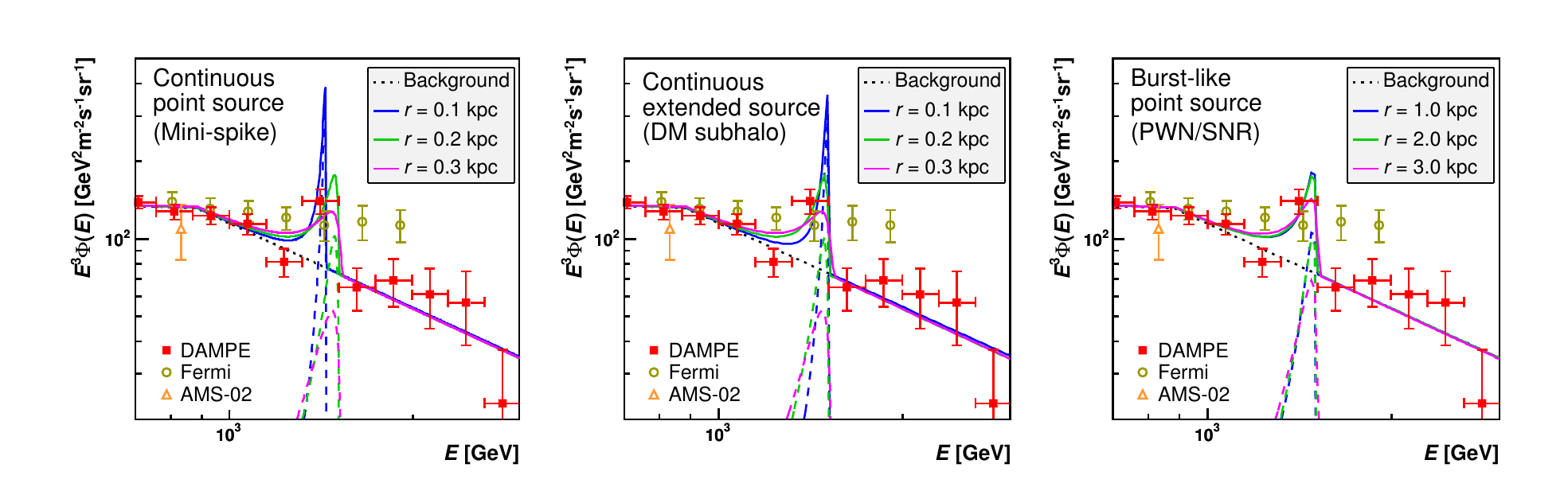}
	\caption{
		Best-fit CRE flux from fitting to the DAMPE data~\cite{Chang:2017xx}  
		for three type of sources.
		Left) Continuous point-like sources (mini-spikes) with distance $r$=0.1, 0.2, 0.3~kpc, 				respectively.
		Center) Continuous extended sources (DM subhalos) for the same distances with subhalo
		mass fixed at $10^{7}M_{\odot}$.
		Right) Burst-like sources (PWNe/SNRs) with $(r (\tx{kpc}),\alpha)$ values 
		(1, 0.5), 
		(2, 0.7), 
		(3, 1.3),			
		respectively.	
		The solid (dashed) curves are  the sum of the signal and background (signal only).
		The data of  DAMPE~\cite{Chang:2017xx},
		AMS-02~\cite{Aguilar:2014fea} and
		Ferm-LAT~\cite{Abdollahi:2017nat} 
		are also shown.
	}\label{fig:flux}  
\end{figure*}

\paragraph{\bf Continuous  sources.}
The propagation of  CR electrons  is described by  
the following diffusion equation
\cite{
	Ginzburg%
}
\begin{align}\label{eq:propagation}
\frac{\partial f}{\partial t}  =
\frac{D(E)}{r^{2}}\frac{\partial}{\partial r}r^{2}\frac{\partial}{\partial r} f
+\frac{\partial}{\partial E}\left(B(E)f\right)
+Q(r,t,E),
\end{align}
where $f(r,t,E)$ is the number density function per unit energy
with $r$ the distance to the source,
$Q(r,t,E)$ is the source term.
In the  equation we have neglected the effects of 
convection and re-acceleration
as they are only important at low energies. %
The energy-dependent spatial diffusion coefficient $D(E)$  is parametrized as
$D(E)=D_{0}(E/\mbox{GeV})^{\delta}$,
where $\delta=0.31$ is a power law index and  
$D_{0}=5.49\times 10^{28}\mbox{ cm}^{2}\mbox{s}^{-1}$ %
\cite{
	Trotta:2010mx%
}. 
The energy-losses due to ICS processes and synchrotron radiations  are prametrized as 
$B(E)=b_{0} E^{2}$ with
$b_{0}=1.4\times10^{-16}~\mbox{GeV}^{-1}\mbox{s}^{-1}$
\cite{
	Linden:2013mqa%
}.

The sources of CRE can be roughly divided into 
continuous  and burst-like sources,
according to the time scale of  electron injection from the sources relative to that of  the propagation time.
A possible continuous source  is the annihilation of DM particles in the Galaxy.
Very sharp $\delta$-function-like CRE spectrum can be produced from
DM annihilation directly into  $e^{+}e^{-}$ pairs 
\note{in models with  enhanced DM-electron coupling,
or  through light mediators with mass very close
to twice the electron mass.
}
However, for any continuous source, the finally observed spectrum is
a superposition of electrons injected at different time.
The electrons injected earlier suffer from more energy losses.
Thus the superposition inevitably results in the broadening of  the spectrum.
For a continuous point-like source
with a $\delta$-function injection spectrum
$Q(r,E) \approx Q_{0}\delta(E-E_{0})\delta^{(3)}(\mathbf{r})$
with $E_{0}$ the central energy and  $Q_{0}$ the normalization constant,
the analytic solution to  \eq{eq:propagation} is  given by
\cite{
	Atoian:1995ux%
}
\begin{align}\label{eq:continuous-point-solution}
f(r,E)=\frac{Q_{0} E^{-2}}{\pi^{3/2}b_{0} 
	r^{3}_{d}(E)}\exp\left(-\frac{r^{2}}{r^{2}_{d}(E)}\right)  ,
\end{align}
where 
$r^{2}_{d}(E)=4 D_{0}
[(E/\mbox{GeV})^{\delta-1}-(E_{0}/\mbox{GeV})^{\delta-1}]/(1-\delta) (b_{0}\text{GeV})$ 
is the diffusion length.
As the solution shows,  after the propagation, 
the  spectrum is broadened.
For $E\ll E_{0}$, it is  
an approximate power law  $f\propto E^{-(1+3\delta)/2}$.
The spectrum rises rapidly when $E$ is approaching $E_{0}$ and 
eventually  cut off exponentially at $E_{0}$ as $r_{d}(E)\approx 0$.
In the left panel of \fig{fig:spectral-feature}, we show how the 
spectral shape of CRE changes with growing distance $r$. 
In the region near  $E\approx E_{0}$  
the spectral shape is very sensitive to the distance.
Increasing the  distance will  result in a broader excess. 
Therefore,  a precision measurement on the spectral shape can be used to 
determine the distance. 
Note that the diffusion length $r_{d}$ 
can only set the scale of the maximal distance.

In the left panel of \fig{fig:flux},
we show the best-fit fluxes obtained from fitting to the DAMPE data 
for three fixed values of  $r=0.1-0.3$~kpc. 
Other parameters such as $E_{0}$, $Q_{0}$ and the background parameters
are allowed to vary in the fits.
In all the  three cases, the best-fit values of $E_{0}$ are quite similar 
$E_{0}\approx1.4-1.5$~TeV. 
With increasing value of $r$, the best-fit spectrum becomes broader and  
the fit  quality becomes lower.
From $r=0.1$ to 0.3~kpc, the $\chi^{2}$-value increases from  $14.2$ to $19.2$.
The fit including  $r$ as a free parameter shows  that 
the DAMPE data place an upper limit of 
$r \lesssim 0.3$~kpc at $95\%$~C.L.. 
For the three cases, the best-fit values of the normalization constants are $Q_{0}$=$(0.47-2.1)\times 10^{33}~\mbox{s}^{-1}$, respectively.
The detailed list of best-fit parameters and allowed regions are shown 
in Tab.~S-2 and Fig.~S-3
of the supplementary material. 

\paragraph{Mini-spikes.}
One of the possible continuous point sources is the ``mini-spike", 
i.e., the large DM density enhancements around the
intermediate mass black holes (IMBHs) with mass $\sim 10^{2}-10^{6} M_{\odot}$
\cite{
	Miller:2003sc,%
	Zhao:2005zr,%
	Bertone:2005xz,%
	Bertone:2009kj%
}.
The IMBH can form 
out of popIII stars
\cite{
	Heger:2002by%
} 
or collapsing of primordial gas in early-forming halos 
\cite{
	Koushiappas:2003zn%
}.
In this letter we consider the latter case of IMBH formation.
For the Milky Way-sized Galaxy, 
the total number of this type of IMBHs is around $\mathcal{O}(100)$
with $\sim 30\%$ of them located in the inner region $ \lesssim10$~kpc
\cite{
	Bertone:2005xz%
}.
Starting from an initial NFW DM profile~\cite{Navarro:1996gj}, 
the spiked DM profile of the mini-spike after the adiabatic growth
of the IMBH follows a power law
$\rho_{sp}(r)=\rho(r_{sp})(r/r_{sp})^{-\gamma_{sp}}$
where 
$\rho(r)$ is the initial DM profile,
$r_{sp}\sim \text{pc}$ is the typical radius of the mini-spike and 
$\gamma_{sp}\approx 7/3$ is the power index%
~\cite{
	Gondolo:1999ef%
}.
Due to the DM annihilation, the spiked DM profile is cut off at a very small distance  $r_{\text{cut}}\sim 10^{-3}$~pc
\cite{
	Bertone:2005xz%
}.
Assuming Majorana DM particles $\chi_{s}$ which annihilate dominantly 
into $e^{+}e^{-}$ pairs  with velocity-averaged  annihilation cross section 
$\langle \sigma v \rangle$, 
the source term of ``mini-spikes" can be estimated as 
\begin{align}\label{eq:Q0}
Q_{0}\approx
&3.1\times 10^{33}~\mbox{s}^{-1}
\left( \frac{\langle\sigma v\rangle}{3\cdot 10^{-26}~\mbox{cm}^{3}\mbox{s}^{-1}} \right) 
\left( \frac{1.4~\mbox{TeV}}{m_{\chi}} \right)^{2}
\nonumber\\
&\left(\frac{\rho(r_{sp})}{10^{2}~\mbox{GeV}\mbox{cm}^{-3}} \right)^{2}
\left(\frac{r_{sp}}{\mbox{pc}} \right)^{14/3}
\left( \frac{r_{\text{cut}}}{10^{-3}~\mbox{pc}}\right)^{-5/3}     .
\end{align}
For the three cases of 
$r=0.1-0.3$~kpc, using the best-fit values of $Q_{0}$, the corresponding cross section are
$\langle \sigma v \rangle=(0.48-2.48)\times 10^{-26}\text{cm}^{3}\text{s}^{-1}$, 
which are close to the typical thermal value.

\paragraph{DM subhalos.}
It is straightforward to extend the analysis to 
the  spatially-extended sources.
N-body simulations of cold and collisionless  DM predict that 
the Galaxy  should contain large number of small subhalos
~\cite{
	Springel:2008cc,%
	Diemand:2008in,%
	Garrison-Kimmel:2013eoa%
}.
\note{
Based on a joint analysis to the Via Lactea II and  ELVIS simulations
\cite{Hooper:2016cld}, 
we estimate that the possibility of finding a nearby subhalo  
within $\lesssim 1 (0.3)$~kpc and total mass 
$M_{h} \gtrsim10^{6} M_{\odot}$ is around 
$\sim 1.2\% (0.03\%)$. 
An alternative possibility is  the ultra compact mini-halos (UCMHs) formed  in the early epochs of the Universe~\cite{
Ricotti:2009bs,
Scott:2009tu,
Josan:2010vn,
Bringmann:2011ut
}.
Finding a nearby UCMH within $\sim0.1$~kpc requires
that UCMHs contribute to a fraction of  above $\sim1\%$~ of the total DM density~\cite{Scott:2009tu}.
In this work, we shall focus on the case of  DM subhalo. 
}
As subhalos may experience a significant degree of mass loss due to tidal stripping,
especially for those located at  the inner volume of the Galaxy,
we  adopt a tidally truncated  density profile
$\rho(r)=\rho_{0}(r/\text{kpc})^{-\gamma}\exp(-r/R_{b})$
\cite{
	Kazantzidis:2003hb,%
	Penarrubia:2007zx,%
	Hooper:2016cld%
}.
The parameters  $\rho_{0}$, $\gamma$ and $R_{b}$  depend on 
the distance  $d$ from the center of the subhalo to the Galactic center
and 
the total mass $M_{h}$ of the subhalo, which  
can be extrapolated from the N-body simulation data.
From the analysis in Ref.
\cite{
	Hooper:2016cld%
},
we obtain $\rho_{0}=5.3~\text{GeV}\cdot\text{cm}^{-3}$, $\gamma=0.78$ and $R_{b}=0.096$~kpc, 
for a typical $M_{h}=10^{7} M_{\odot}$.

The CRE flux from the subhalo is calculated by integrating the solution of 
\eq{eq:continuous-point-solution} over the subhalo density distribution.
For $r=0.1$, 0.2, and 0.3~kpc,
we find  the best-fit annihilation cross sections 
$(1.04, \ 2.55,\  4.62)\times 10^{-26}~\mbox{cm}^{3}\text{s}^{-1}$, respectively,
which are well below  the current limits on DM subhalos
\cite{
	Fermi-LAT:2016uux,%
	Schoonenberg:2016aml,%
	Hooper:2016cld%
}.
Similar to the case of point source, 
when $r$ increases from 0.1 to 0.3~kpc, 
the best-fit spectrum becomes broader and the fit qualities become lower.
The corresponding $\chi^{2}$ value increases from 13.4 to  19.2.
From a fit with $r$ as a free parameter, 
we find that  the source should be within $r \lesssim 0.3$~kpc at $95\%$~C.L., 
very close to the case of mini-spikes.
We also find that modifying the subhalo mass $M_{h}$ 
does not change the conclusion.
The detailed list of best-fit parameters and allowed regions
are shown in Tab.~S-4 and Fig.~S-4
of the supplementary material. 
Note that 
for  an  annihilation cross section of 
$\mathcal{O}(10^{-26})~\tx{cm}^{3}\tx{s}^{-1}$ %
the contribution from the whole Galactic halo DM 
to the CRE flux  is typically two orders of magnitude  smaller than 
that from the nearby sources,
which can be safely neglected.
\note{
The halo DM also contribute to extra CR positrons which 
could be constrained by the experiments.
The constraints, however, turn out to be rather weak for 
TeV scale DM, 
as the current AMS-02  experiment only measured the 
positrions up to 
$\sim 350$~GeV~\cite{Accardo:2014lma}
(see Fig.~S-6 in the supplementary material).
}

\paragraph{\bf Burst-like  sources.}

For burst-like sources such as PWNe/SNRs,  
the injection spectrum is expected to be
a broad power-law with an exponential cutoff,
$Q(r,t,E)=N_{0} (E/\mbox{GeV})^{-\alpha} 
\exp(-E/E_{c}) \delta^{(3)}(\mathbf{r})\delta(t)$, 
where $\alpha$ is the power-law index and
$E_{c}$ is the cutoff energy.
The normalization constant $N_{0}$ can be  related to 
the total released energy  $E_{\tx{tot}}$.
The solution to the diffusion equation~\eq{eq:propagation} 
for this type of source is given by~\cite{Atoian:1995ux}
\begin{align}\label{eq:burst--solution}
f(r,E)=\frac{N_{0}
	(E/\text{GeV})^{-\alpha}}{\pi^{3/2}r_{d}^{3}} 
\xi(E)^{\alpha-2}
\exp\left(-\frac{r^{2}}{r_{d}^{2}}-\frac{E}{\xi(E) E_{c}}\right) ,
\end{align}
where $\xi(E)=1-E/E_{\text{max}}$ 
with $E_{\text{max}}=(b_{0}t)^{-1}$  the maximal possible energy of 
an electron from a source of age~$t$.
The diffusion length for this type of source is
$r_{d}(E) =2\sqrt{ \lambda(E) D(E)t}$,
where $\lambda(E)=[1-\xi(E)^{1-\delta}]/(1-\delta)(1-\xi(E))$.
For $E\ll E_{\text{max}}$, $r_{d}(E) \approx 2\sqrt{D(E)t}$.
While the value of $\alpha$ is commonly considered to be $\sim 2$,
the cutoff $E_{c}$ is poorly constrained.
We shall focus on the large cutoff limit, 
i.e. $E_{c}\gg E_{\text{max}}$.
In this case,
$E_{\text{max}}$ will play the role of spectral cutoff 
instead of $E_{c}$, namely,
a cooling cutoff will appear,
as can be seen in \eq{eq:burst--solution}.

In this work, we emphasize that for the burst-like source 
a sharp spectral rise near the cutoff energy $E_{\text{max}}$
can appear for some choices of $\alpha$ and $r$
for two reasons:
{\bf i)}  Cooling related ``phase-space shrinking''.
An initial  electron with energy $E_{s}$ at time $t=0$
is related to its energy $E$ observed at later time $t$ as $E_{s}=E/\xi(E)$.
Thus an initial energy interval  $\Delta E_{s}$ will shrink to 
$\Delta E=\xi(E)^{2} \Delta E_{s}$ at time $t$.
Since the number of electrons  is unchanged during cooling,
the energy spectrum at time $t$ is  
$\Delta N/\Delta E \approx E^{-\alpha}\xi(E)^{\alpha-2}$.
For a relatively hard spectrum with  $\alpha <2$, 
the shrinking of phase space  can enhance the number density.
Since $\xi(E)$ vanishes when $E$ is approaching $E_{\text{max}}$,
the shrinking of phase space leads 
to a rapid rise of the energy spectrum.
This effect of ``phase-space shrinking'' is illustrated  in
the right panel of \fig{fig:spectral-feature}.
The DAMPE data suggest that the cutoff  should be 
in the range $E_{\text{max}}\approx 1.3-1.5$~TeV,
which in turn sets the  age of the source
$t  %
\approx  (0.15-0.17)~\text{Myr}$,
and the diffusion length at $E_{\text{max}}$, 
$r_{d}(E_{\text{max}}) \approx (0.79-0.84)~\text{kpc}$.
{\bf ii)}``Energy-dependent diffusion'' which is related to 
the fact that for $\delta>0$,
higher energy electrons have larger diffusion coefficients.
The energy dependence in the exponential factor
of \eq{eq:burst--solution} can be written as
$\exp(-r^2/r_d^2)\approx \exp[- \kappa^{2}(E/E_{\text{max}})^{-\delta}]$
where $\kappa=r^{2}/r_{d}^{2}(E_{\text{max}})$.
For relatively  large distance  $\kappa >1$, 
the energy-dependent factor also contributes to the rising of the spectrum near $E_{\text{max}}$,
which  is illustrated in  \fig{fig:spectral-feature}.
Of course, in order to  compensate the exponential suppression of the flux at 
large $\kappa$, 
the  normalization constant $N_{0}$ 
or $E_{\text{tot}}$ has to be large enough.
The reasonable value of $E_{\text{tot}}$ should be smaller than 
the typical kinetic energy carried by SNR 
or the total energy of supernova explosion of  
$\sim(10^{51}-10^{53})$~erg,
which sets the scale of the distance of the sources.

In general, unconventional values of  $\alpha$ 
which is  significantly smaller than $\sim 2$ is required to 
reproduce the DAMPE excess, 
especially for small $r$.
In the right panel of  \fig{fig:flux}  we show the best-fit spectra for  
three typical combinations of  $r$ (in kpc) and $\alpha$ with 
$(r,\alpha)$=(1, 0.5), (2, 0.7) and (3, 1.3), respectively.
A scan in the $(r,\alpha)$ parameter space shows that in the region 
$r<r_{d}$, the allowed $\alpha$ has to be very small $\alpha \lesssim0.65$.
In the region $r>r_{d}$ the value of $\alpha$ can reach at most $1.4$ at 3~kpc,
as the effect of  energy-dependent diffusion is  significant.
By imposing the condition of $E_{\text{tot}}<10^{51}(10^{53})$~erg,
the distance $r$ is restricted in the range $r\lesssim$3(4)~kpc.
Together with the required  age of the source,
we find 7 candidate pulsars with $r \lesssim 4$~kpc
in the ATNF catalog of pulsars~\cite{Manchester:2004bp}:
B0740-28, J0922-4949,   J1055-6022,  J1151-6108,
J1509-5850,  J1616-5017 and J1739-3023.
Distances of the sources lie in the range (2.0--3.6)~kpc.
In this region, both the effects of ``phase-space shrinking '' and 
``energy-dependent diffusion'' are relavant.
The detailed list of best-fit parameters, allowed regions
and the list of  the candidate pulsars are shown in
Tab.~S-6, Fig.~S-7 and Tab.~S-7
of the supplementary material. 
\note{
For both continuous and burst-like sources, 
varying the propagation parameters
$D_{0}$ and $\delta$ within uncertainties 
($\sim 20\%$ for $D_{0}$ and $\sim10\%$ for $\delta$
as determined in \cite{Trotta:2010mx}) mainly results in 
the changes in the over all normalization factors 
$Q_{0}$ and $N_{0}$ up to $\sim30\%$.
}

\begin{figure}[!thbp]
	\begin{center}
		\includegraphics[width=0.65\columnwidth]{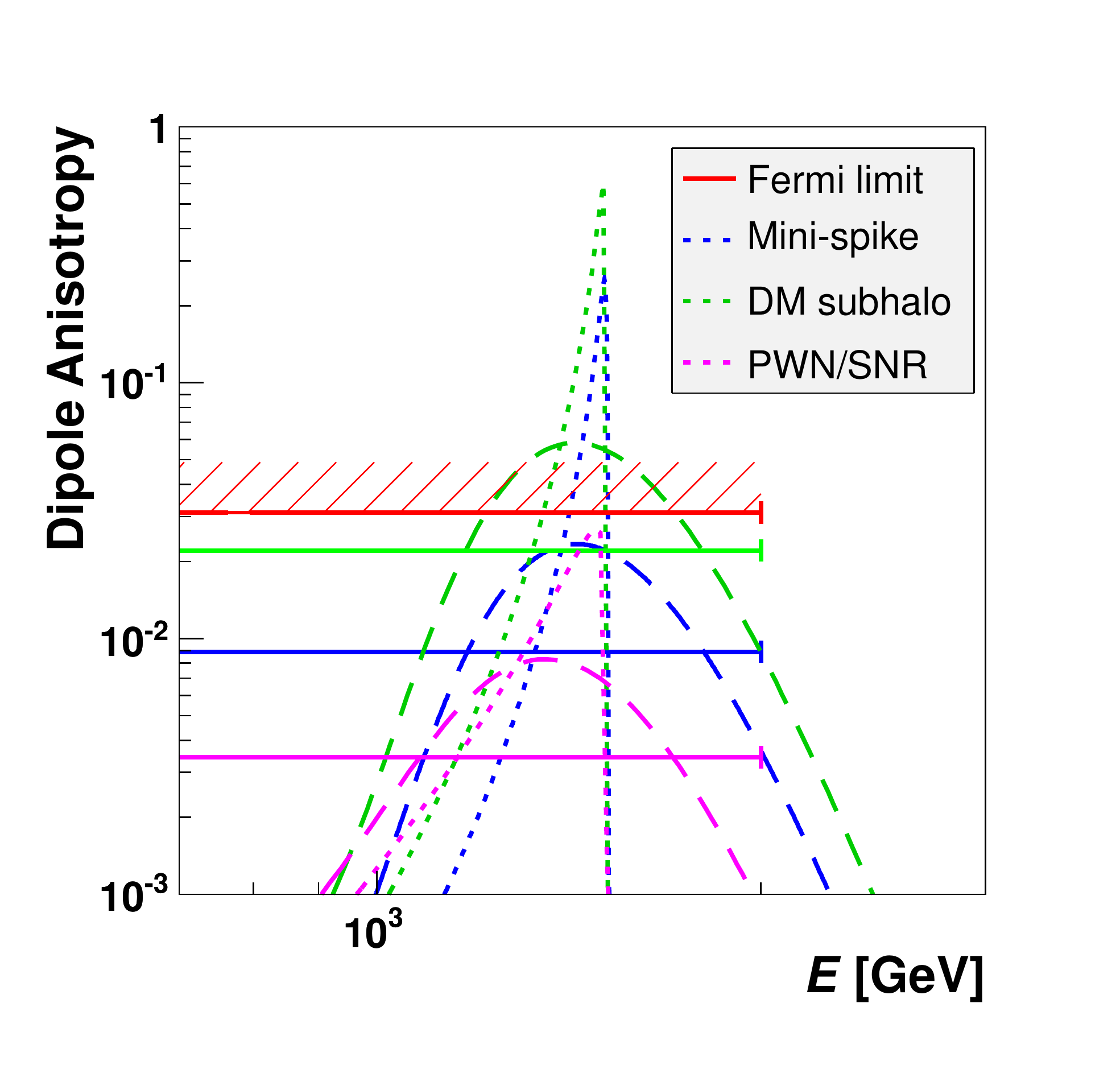}
		\caption{
			Predictions for electron anisotropies corresponding to 
			a selection of the cases considered in \fig{fig:flux}:
			i) the continuous point-like source (mini-spike) with $r=0.2$~kpc (dashed blue);
			ii) the continuous extended source (subhalo) with $r=0.2$~kpc and $M_{h}=10^{7}M_{\odot}$ (dashed green);
			iii) the burst-like source (PWN/SNR) with $r=2$~kpc and $\alpha$=0.7 (dashed magenta).
			The solid curves with the same color correspond to the anisotropies convoluted with 
			an energy resolution of $15\%$.
			The current upper limits from Fermi-LAT in the  energy interval 0.55--2~TeV
			(using the method of shuffling technique)
			are also shown%
			~\cite{Abdollahi:2017kyf}.
		}
		\label{fig:anisotropy-combined}
	\end{center}
\end{figure}

\paragraph{\bf Anisotropies.}
Nearby sources can generate non-negligible anisotropy in the CRE flux.
For an illustration,
we show in \fig{fig:anisotropy-combined}  the predicted 
dipole anisotropies from the sources as a function of the CRE energy,
corresponding to one of the  parameter sets
considered in each type of the sources shown in \fig{fig:flux}.
Assuming  a perfect energy resolution of the detector,
large anisotropies of $\mathcal{O}(10^{-1})$ with 
sharp structures  are predicted.
The anisotropies in continuous sources are in general 
larger than that in the burst-like sources,
which is  related to the relatively small diffusion length.
The Fermi-LAT  has reported  upper limits on the  
dipole anisotropy of $\lesssim3\times 10^{-2}$  at $95\%$ C.L.
over the energy interval  0.55--2~TeV%
~\cite{Abdollahi:2017kyf}.
Note that the energy resolution of Fermi-LAT is $\sim 10\%~(17\%)$ 
at 1 (2)~TeV.
In  \fig{fig:anisotropy-combined}, 
we also show  the predicted anisotropies convoluted 
with an energy resolution of $15\%$.
After the convolution, 
the predicted anisotropies are  smaller and can reach $\mathcal{O}(10^{-2})$,
which is comparable with  the current Fermi-LAT limits.
Note that a quantitative comparison with the data requires 
a reliable estimation of the anisotropies contributed by the backgrounds 
which can easily reach $\mathcal{O}(10^{-3}-10^{-2})$ alone,  but 
depends strongly on the assumed spatial distribution of 
the astrophysical sources%
~\cite{Manconi:2016byt}.
%


\begin{figure}[thb]
	\begin{center}
		\includegraphics[width=0.49\columnwidth]{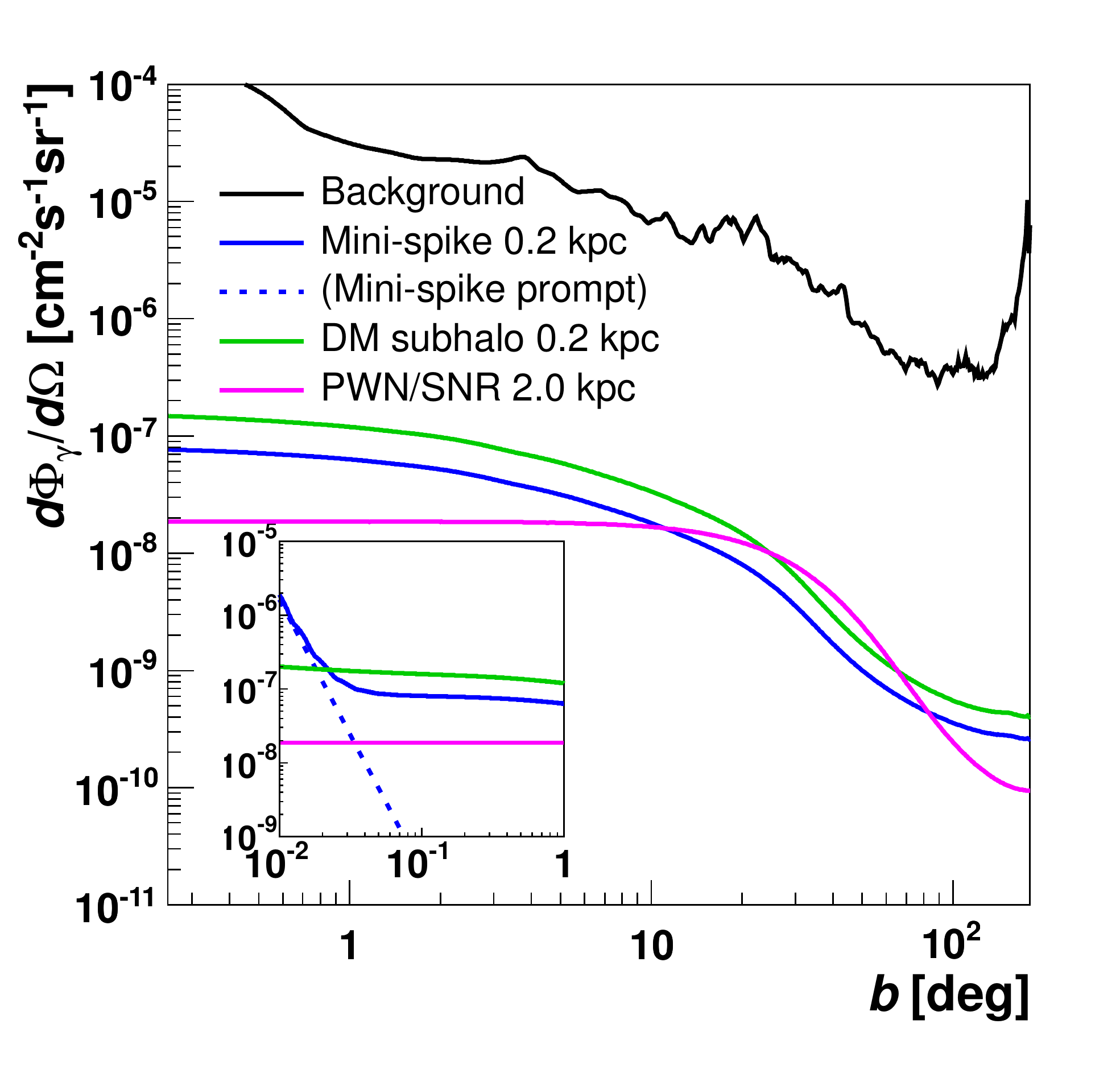}
		\includegraphics[width=0.49\columnwidth]{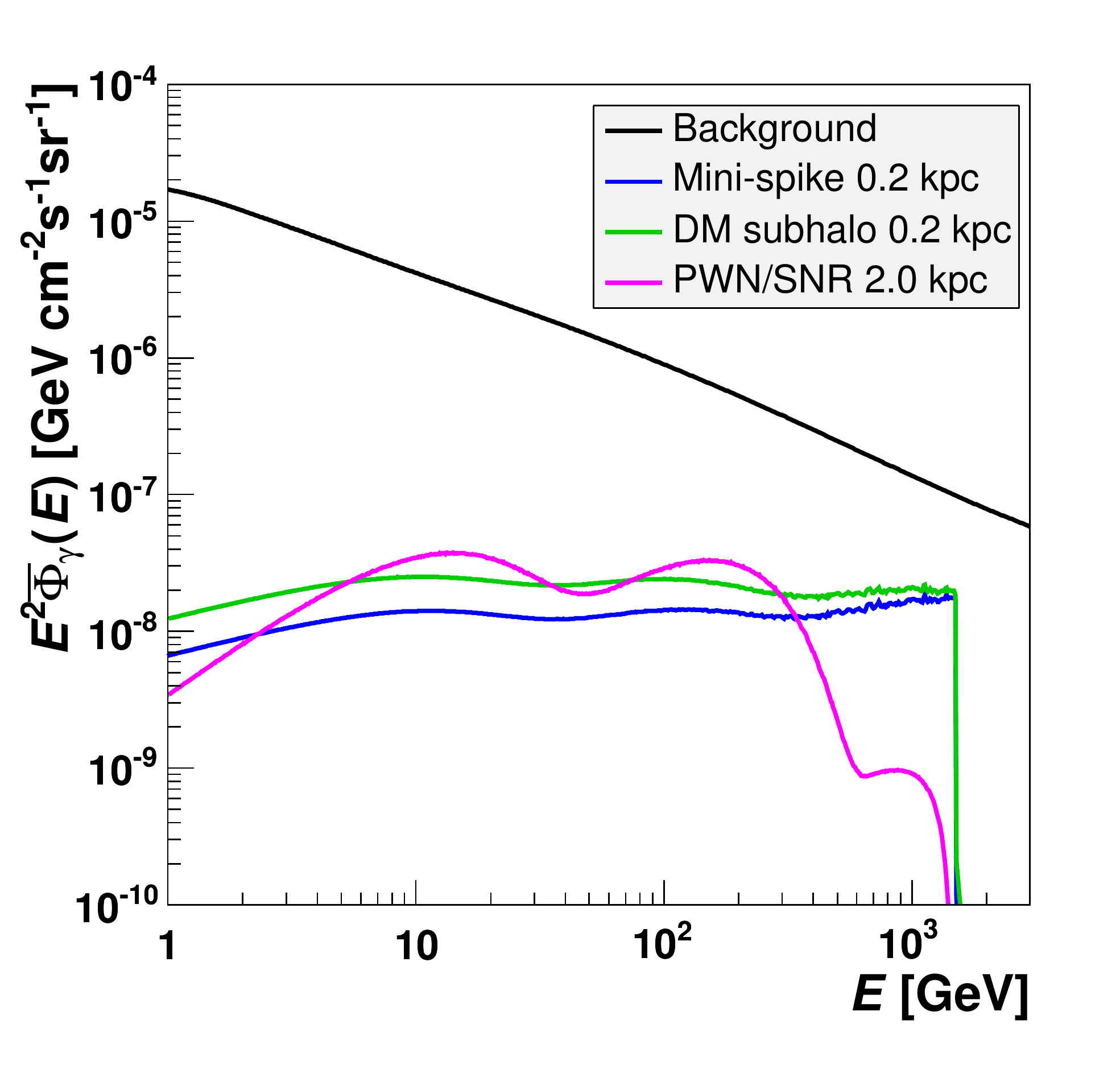}
		\caption{
			Predictions for the Galactic diffuse $\gamma$-ray fluxes in the three cases
			considered in \fig{fig:anisotropy-combined} assuming the direction of the sources
			is coincide with the GC. 
			Left) 
			$d\Phi_{\gamma}/d\Omega$ as a function of Galactic latitude $b$
			along the direction of  $\ell=0^{\circ}$.
			The inset shows the fluxes  in the  inner region,
			which indicates that the $\gamma$-rays from the mini-spike
			increase sharply towards low $b$ due to prompt photons from DM annihilation, 
			making it appears as a point-like source.
			Right)
			Energy spectra of $\gamma$-ray fluxes averaged over a circular region with radius $30^{\circ}$
			centered at the GC.
			The Galactic diffuse $\gamma$-ray background is 
			calculated using  GALPROP-v54 \cite{galprop}
			with a reference propagation model 
			adopted by Fermi-LAT
			\cite{
				Ackermann:2012pya%
			}.
			The ISRF in the solar neighbourhood is interpolated from Ref.
			\cite{
				Porter:2008ve%
			}.
		}
		\label{fig:GAMMA}
	\end{center}
\end{figure}

\paragraph{\bf Gamma-ray signals.}
DM annihilation can produce prompt photons through final state radiation (FSR)  of charged leptons. 
The FSR photon spectrum has a distinct feature of an approximate power law with index $\sim1$ for $E\ll m_{\chi}$, plus a sharp cutoff at $m_{\chi}$.
In morphology, 
mini-spikes appear as point-like sources due to 
the high concentration of DM density.
For the cases of mini-spike considered in \fig{fig:flux}, 
the total fluxes ($E>1$~GeV) 
are in the range  
$\sim (0.71-1.51)\times 10^{-10}\text{cm}^{-2}\text{s}^{-1}$.
In the Fermi-LAT 3FGL catalogue of unassociated point sources~\cite{Acero:2015hja},
we find 6 candidate  sources with low power indices
which can reach $\sim1$ within $2\sigma$ error:
J0603.3+2042, J1250.2-0233, J2209.8-0450, J1705.5-4128c, J2142.6-2029
and J2300.0+4053.
Most of them have total fluxes of 
$\mathcal{O}(10^{-10})~\text{cm}^{-2}\text{s}^{-1}$  %
which are consistent with that favoured by the mini-spikes.
The favoured 
DM annihilation cross sections 
for both mini-spikes and DM subhalos
are well below the  current upper limits derived from  
the $\gamma$ rays towards 
dwarf Galaxies~\cite{Ackermann:2015zua} and 
the Galactic center (GC)~\cite{Abdallah:2016ygi}
under the assumption of   smooth DM  profiles.
%
Taking into account the distribution of mini-spikes,
very stringent constraints on the annihilation cross sections were
obtained by HESS from the GC $\gamma$-ray data 
\cite{%
	Aharonian:2008wt%
}, 
which are highly model dependent. 
%
The $\gamma$-rays constraints   from 
other sky regions are much weaker
~\cite{Bringmann:2009ip}.
%

All the CRE sources can produce diffuse $\gamma$-rays through 
electron inverse Compton scattering (ICS) off 
interstellar radiation fields (ISRF),
the corresponding energy spectrum is softer compared with 
that of the FSR photons~\cite{Meade:2009iu}.
For the typical parameters considered in the three scenarios in \fig{fig:flux}, 
the predicted differential flux $E^{3}d\Phi/dE$ can reach 
$\mathcal{O}(10^{-8})~\tx{GeV}\tx{cm}^{-2}\tx{s}^{-1}\tx{sr}^{-1}$ 
in the energy range $\sim(0.1-1)$~TeV,
which may be subject to the constraints from the Fermi-LAT data 
on the Galactic diffuse $\gamma$-rays~\cite{Ackermann:2012rg}.
However, the Fermi-LAT constraints vary with sky regions.
In the  low Galactic latitudes regions the backgrounds can easily 
reach $\sim10^{-6}~\tx{GeV}\tx{cm}^{-2}\tx{s}^{-1}\tx{sr}^{-1}$
and dominate the total diffuse $\gamma$-ray  emission. 
In this case the  Fermi-LAT data cannot place stringent constraints 
on these CRE  sources.
This possibility is illustrated in \fig{fig:GAMMA}, where
the direction of the source is assumed to  coincide with the GC.
For the three cases discussed  in  \fig{fig:anisotropy-combined},
we show the spatial extension as well as the energy spectrum of the associated $\gamma$-rays,
together with the corresponding Galactic diffuse $\gamma$-ray backgrounds.
In the calculations, 
the contributions from the Galactic halo DM 
for mini-spikes and DM subhalos are included
assuming an Einasto DM profile.
The Galactic diffuse $\gamma$-ray background is 
calculated using a reference propagation model 
adopted by Fermi-LAT~\cite{Ackermann:2012pya}
which agrees with the data well.
It can be seen that for  the typical parameters  considered, 
the predicted $\gamma$-ray fluxes can be 
a few orders of magnitude below the background,
which suggests that the Ferm-LAT constraints should be 
rather weak, 
as the uncertainties in the background model are still significant
(see the supplementary material for detailed calculations on 
the diffuse $\gamma$-rays).
%


\begin{acknowledgments}
This work is supported in part by the National Key R$\&$D Program of China 
No. 2017YFA0402204, the NSFC under Grants No. 11335012, 11690022, 11475237 and U1738209,
and the CAS Key Research Programs, No. XDB23030100 and QYZDY-SSW-SYS007.
{\it Note added.} After submitting the first version of the manuscript,
a number of analyses on the DAMPE data appeared%
~\cite{
Yuan:2017ysv,Fan:2017sor,Fang:2017tvj,Duan:2017pkq,Gu:2017gle,
Athron:2017drj,Cao:2017ydw,Liu:2017rgs,Zu:2017dzm,Tang:2017lfb,
Chao:2017yjg,Gu:2017bdw,Duan:2017qwj,Cholis:2017ccs,Jin:2017qcv,
Gao:2017pym,Niu:2017hqe,Chao:2017emq,Chen:2017tva,Li:2017tmd,
Zhu:2017tvk,Gu:2017lir,Nomura:2017ohi,Ghorbani:2017cey,Cao:2017sju,
Yang:2017cjm,Ding:2017jdr,Liu:2017obm,Ge:2017tkd,Zhao:2017nrt,
Sui:2017qra,Okada:2017pgr,Cao:2017rjr,Dutta:2017sod,Fowlie:2017fya,Han:2017ars,
Niu:2017lts%
}.
We found that most of the proposed models fall into the scenarios 
discussed in our work.
Ref.~\cite{Yuan:2017ysv} discussed the contributions from PWNe and DM substructures.
Their conclusions are also consistent with ours.
\end{acknowledgments}

\bibliographystyle{apsrev4-1} %
\bibliography{dampe,misc,dampe_papers}

\clearpage
\onecolumngrid
\input{supp.tex}

\end{document}

%% file: supp.tex
\section{Supplementary Material}

\renewcommand{\theequation}{S-\arabic{equation}}
\setcounter{equation}{0}

\renewcommand{\thefigure}{S-\arabic{figure}}
\setcounter{figure}{0}

\renewcommand{\thetable}{S-\arabic{table}}
\setcounter{table}{0}

\vskip 12pt
\centerline{\bf \large Origins of  sharp cosmic-ray electron structures and the DAMPE excess}
\centerline{\it Xian-Jun Huang, Yue-Liang Wu, Wei-Hong Zhang, and Yu-Feng Zhou}
\vskip 12pt

In this supplementary material, we give more details and discussions of the results obtained 
in the letter.

\subsection{A. Properties of the DAMPE excess}
\begin{figure}[h]
	\centering
	\includegraphics[width=0.4\columnwidth]{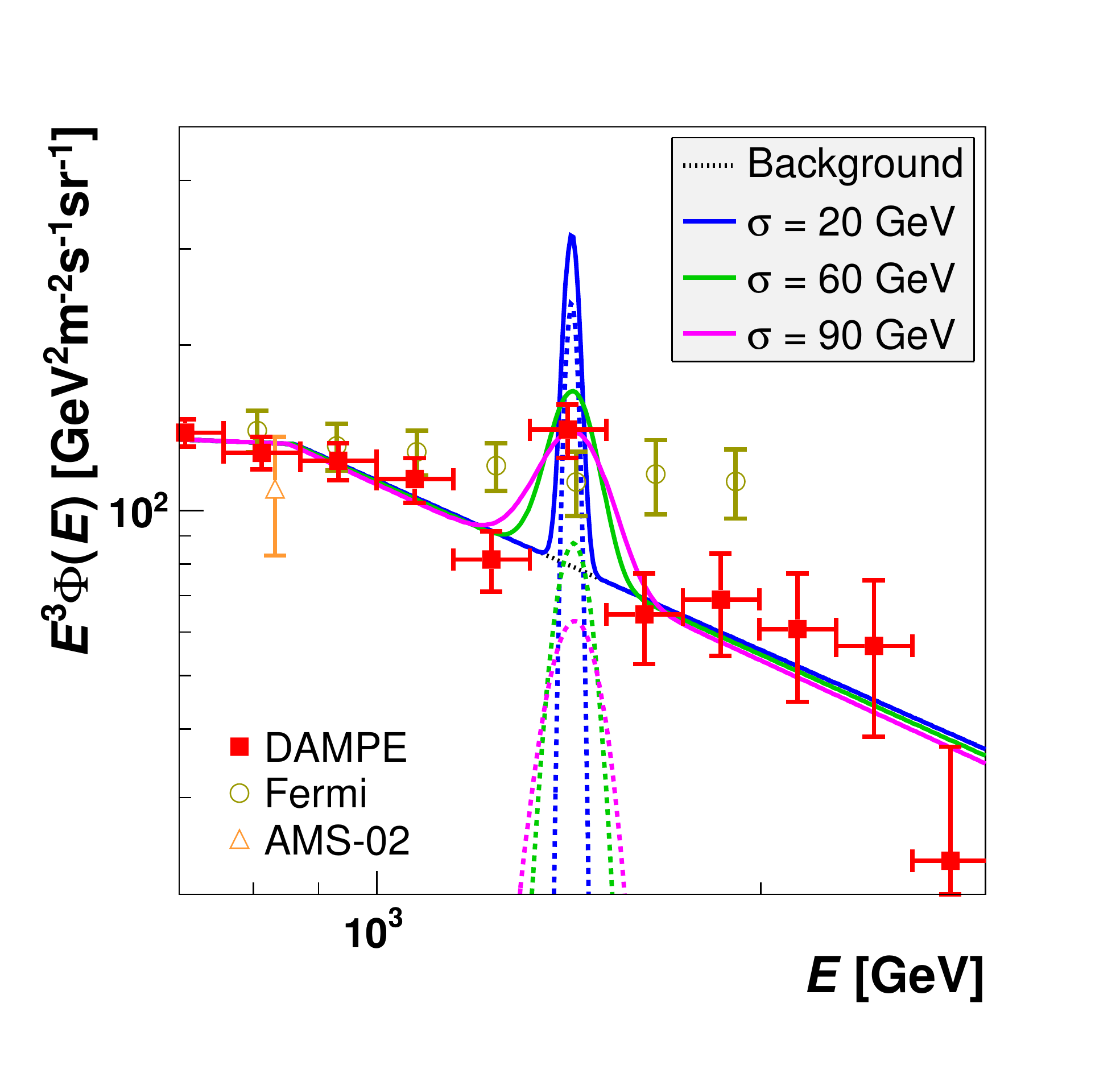}
	\caption{
	Best-fit energy spectra in the scenarios of background-only and
	the  background plus a Gaussian signal with width
	$\sigma=20$, 60 and 90~GeV, respectively.
	The data of  AMS-02 [2], 
	Ferm-LAT [3], 
	and 
	DAMPE [4] 
	are also shown.
	}
	\label{fig:flux_SNR_multi}
\end{figure}

We adopt a two-component description of the 
CRE flux, namely, the main bulk of the CRE flux
arises  from  distant astrophysical sources in the Galaxy,
which  leads to  a smooth power law-like background.
In addition, there is a dominant nearby source of high energy CRE which 
contributes to an excess over the smooth background.
To estimate the significance of the DAMPE excess, 
we consider a  reference  background model in which  the flux $\Phi_{b}$ 
is assumed to be a broken power law 
$\Phi_{b}(E)=N_{b} (E/E_{\tx{brk}})^{-\gamma}$, 
where $\gamma=\gamma_{1(2)}$ for 
the electron energy below (above) a break energy $E_{\text{brk}}$, 
and $N_{b}$ is a normalization factor. 
The flux of  the  excess is parametriezed as  Gaussian 
$\Phi_{s}(E)=N_{s} \exp[(E-\mu)^{2}/2\sigma^{2}]$,
where
$\mu$ and $\sigma$ are the central value and half-width, respectively,
and $N_{s}$ is a normalization factor.
We fit to the DAMPE data in the energy interval $55$~GeV--4.6~TeV.
In total 32 data points are included.
The model parameters are determined through 
minimizing the $\chi^{2}$-function defined as follows
\begin{align}
\chi^{2}=\sum_{i}
\frac{(\Phi_{b,i}+\Phi_{s,i}-\Phi_{\text{exp},i})^{2}}{\sigma_{\text{exp},i}^{2}},
\end{align}
where $\Phi_{b(s),i}$ is the theoretical value of the  
background (excess)  {\it averaged} over the width of the $i$-th energy bin,
$\Phi_{\text{exp},i}$ and $\sigma_{i}$ are the measured mean value 
and error of the flux in the energy bin.
The effect of finite energy resolution is neglected as the energy resolution
of DAMPE is quite high, better than 1.5$\%$ at 800~GeV.
The minimization of $\chi^{2}$ is performed using the MINUIT package.
The fit results in the scenarios of background-only and background plus the 
excess are  summarized in \tab{tab:gaussfree}.
It can be seen that in the background-only scenario
while the power index $\gamma_{1}$ is well constrained 
with uncertainty within $\sim 0.3\%$,
the uncertainty in $\gamma_{2}$ is much  larger around $\sim 5\%$,
which is due to the limited statistics at high energies.
Compared with the background-only scenario, the inclusion of 
a Gaussian excess leads to a reduction of $\Delta\chi^{2}=13.7$, 
roughly corresponding to  $\sim 3.7\sigma$ local significance.
Since  the excess is observed in a single energy bin, 
the lower limit of the width $\sigma$ cannot be determined,
namely, the possibility of an electron-line is not ruled out for the moment.
We find that at $68\% ~(95\%)$~C.L., 
the upper limit on the width is $\sigma \leq 60.4~(89.6)$~GeV.
In \fig{fig:flux_SNR_multi}, we show the best-fit fluxes for fixed
$\sigma=20,\ 60$ and 90~GeV, the corresponding $\chi^{2}$ values are
$12.2$, 13.2 and 16.2, respectively.

\begin{table}[htbp]
	\centering
		\begin{tabular}{c|ccccccc|c}
		\hline\hline
		   & $E_{\text{brk}}$ (GeV) & $\gamma_1$ & $\gamma_2$ & $N_b $ & $\mu$ (GeV) &$\sigma$ (GeV) & $N_s$& $\chi^2/\text{d.o.f.}$ \\
		 \hline
		 background-only & $884^{+93}_{-93}$ & $3.10\pm 0.01$ & $3.95^{+0.19}_{-0.16}$& $1.91^{+0.81}_{-0.52}$  & ...&...&...&25.85/28 \\
		background+excess & $863^{+74}_{-90}$& $3.10 \pm0.01$ & $4.03^{+0.19}_{-0.18}$ & $2.06^{+0.86}_{-0.48}$ & $1418.2^{+95.4}_{-99.9}$ & $0^{+60.36}_{-0.0}$ & $0.643$ &12.18/25 \\
		\hline\hline
		\end{tabular}
	\caption{
	Best-fit parameters in the scenarios of background-only and background plus a 
	Gaussian excess. The normalization constants $N_b, N_s$ are in units of 
	$10^{-7}$ GeV$^{-1}$m$^{-2}$s$^{-1}$sr$^{-1}$.}
\label{tab:gaussfree}
\end{table}

\subsection{B. Continuous sources}

\subsubsection{Point-like sources}
\begin{figure}[htb]
	\centering
	\includegraphics[width=0.4\textwidth]{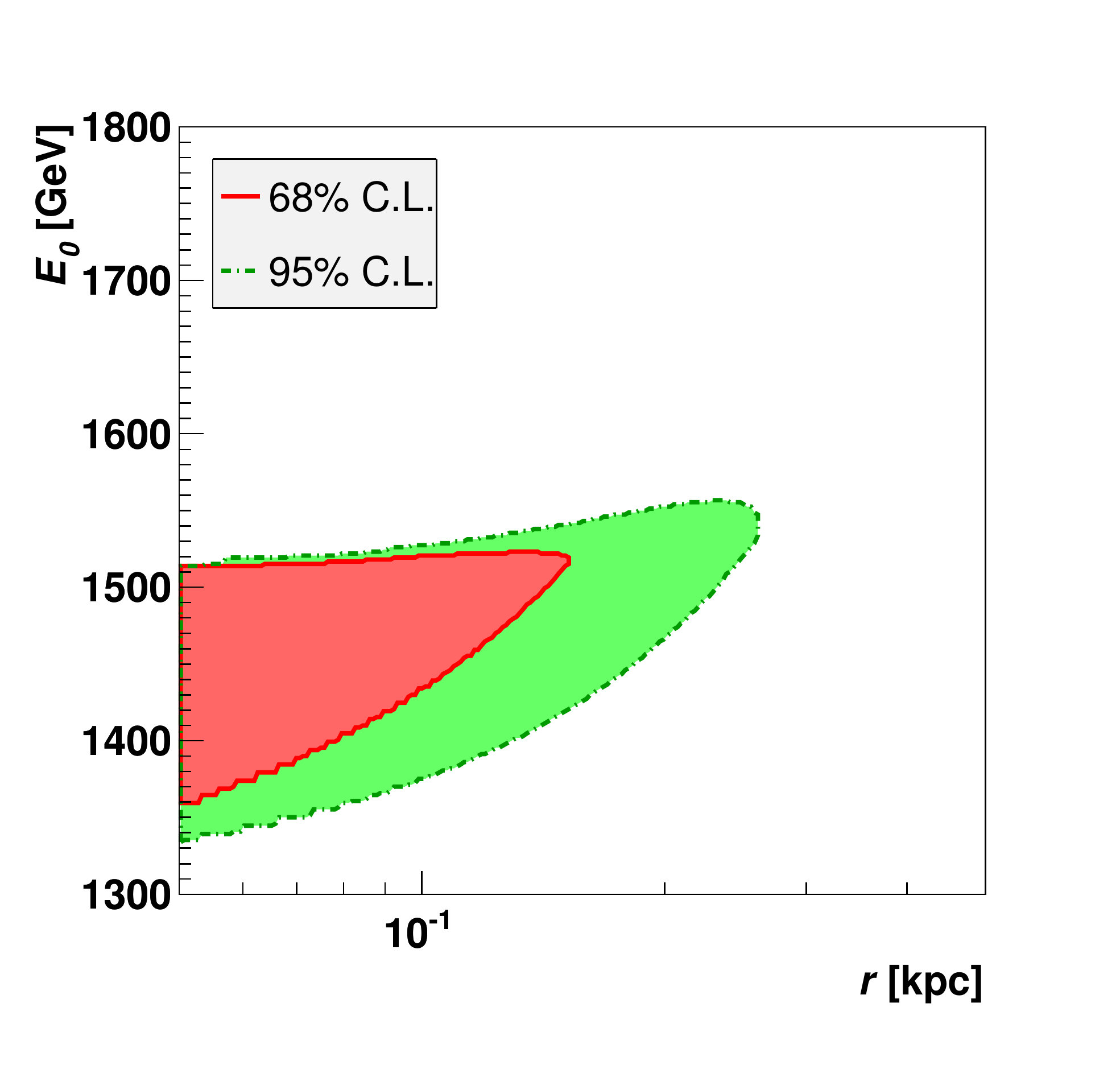}
	\includegraphics[width=0.4\columnwidth]{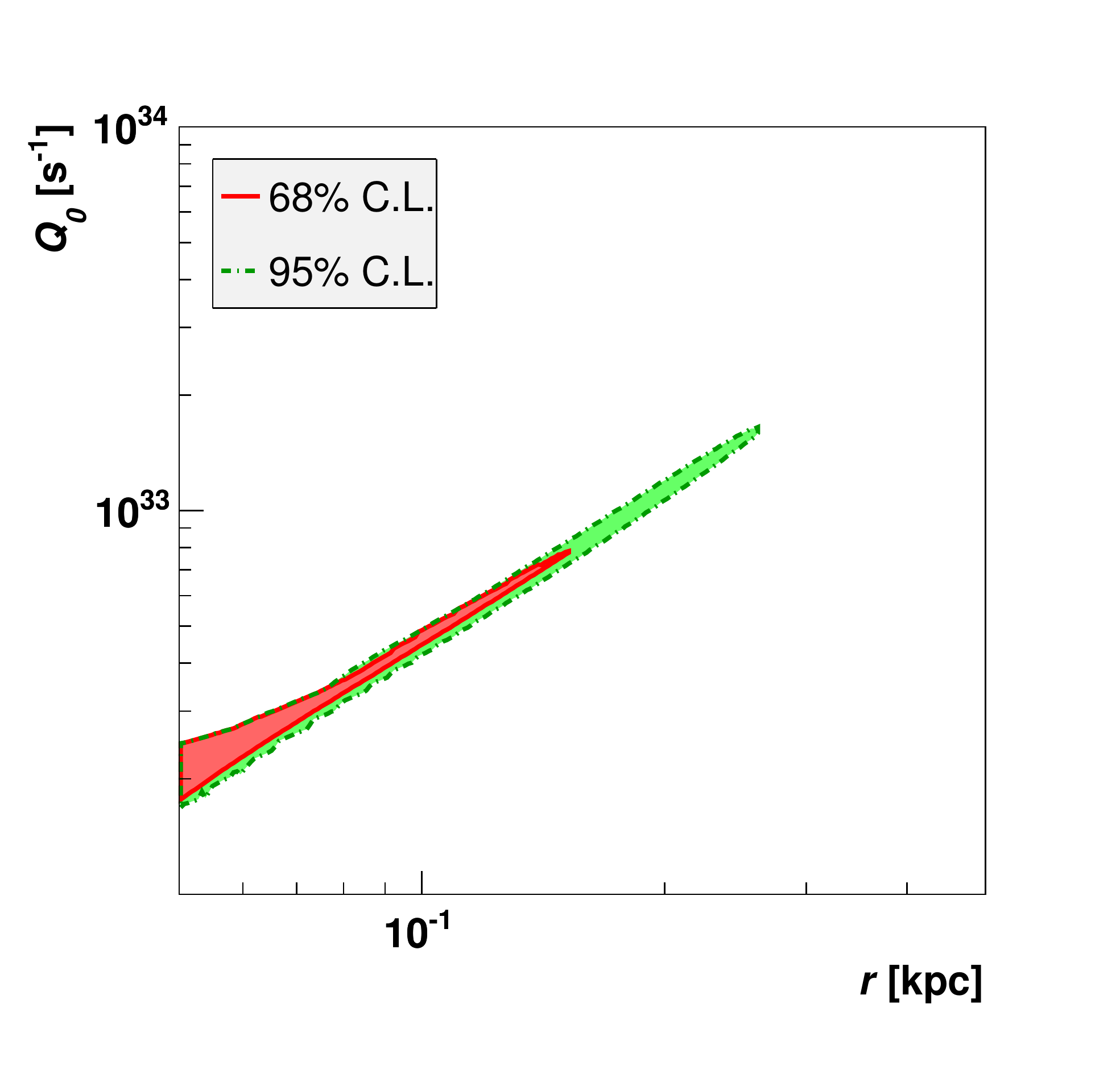}
	\caption{		
	Left) 
	Allowed regions at $68\%$ and $95\%$ C.Ls. in $(r, E_{0})$ plane for 
	continuous point-like sources.
	Right)
	The same as Left, but for the allowed regions in $(r, Q_{0})$ plane.	
	}\label{fig:pointDM-contour}
\end{figure}

The fit results for the scenarios of continuous point-like sources for 
three representative choices of distances $r=0.1$, 0.2 and 0.3~kpc,
together with the best-fit background parameters  are 
summarized in \tab{tab:pointDM},
where the break energy of the background is fixed at 
the best-fit value in the background-only scenario, i.e., $E_{\text{brk}}=884$~GeV.
The propagation parameters $D_{0}$, $\delta$ and $b_{0}$ are given in the letter.
\begin{table}[htbp]
	\centering
		\begin{tabular}{c|ccccc|c}
		\hline\hline
		  $r$ [kpc] & $Q_0\ (10^{32} \text{ s}^{-1})$ & $E_0$ (GeV)& $\gamma_1$ & $\gamma_2$ & $N_b$  & $\chi^2/\text{d.o.f.}$ \\
\hline
0.1 & $4.67^{+1.39}_{-1.48}$& $1446^{+32}_{-77}$& $3.10\pm0.01$ & $4.08^{+0.14}_{-0.13}$& $1.90\pm0.03$  & 14.18/27 \\
0.2 & $11.84^{+3.82}_{-4.00}$ & $1525^{+34}_{-19}$ & $3.10\pm0.01$ & $4.09^{+0.15}_{-0.13}$ & $1.90\pm0.03$ & 15.81/27 \\
0.3 &  $20.78^{+8.28}_{-8.62}$ & $1554^{+34}_{-33}$ & $3.10\pm0.01$ & $4.09^{+0.16}_{-0.14}$ & $1.89^{+0.03}_{-0.04}$ & 19.15/27 \\
\hline\hline
\end{tabular}
	\caption{
	Best-fit parameters  for continuous point-like sources with distance $r$=0.1, 0.2 and 0.3 kpc.
	The normalization constant $N_b$ is in units of  
	$10^{-7}$GeV$^{-1}$m$^{-2}$s$^{-1}$sr$^{-1}$.
	The best-fit values of $Q_{0}$ correspond to the DM annihilation cross section 
	$\sigmav=4.82\times10^{-27}$, $1.36\times10^{-26}$, 
	and $2.48\times10^{-26} \text{cm}^3\text{s}^{-1}$ for
	$r=0.1$, 0.2 and 0.3~kpc, respectively.
	}
\label{tab:pointDM}
\end{table}
In the case where the distance $r$ is allowed to vary freely, 
the value of $\chi^{2}$ decrease with decreasing $r$ and gradually reaches 
a minimal value  $\chi^{2}_{\tx{min}}$=13.0 in the limit  $r\to 0$. 
This result is again related to the fact that the excess is observed in a single
energy bin and the $\chi^{2}$ value is based on the bin-average.
We thus define the  allowed regions in $(r, E_{0})$ and $(r,Q_{0})$ planes  
at $68\%$ ($95\%$)~C.L. for two parameters,
corresponding  to $\Delta \chi^{2}=2.3$ (5.99),
which are shown  in \fig{fig:pointDM-contour}.
It can be seen that the allowed distance is constrained to be
$r \lesssim 0.3$~kpc at $95\%$~C.L.
The value of $E_{0}$ is constrained in the $1.35-1.55$~TeV range which is compatible with the 
width of the excess.

\subsubsection{Mini-spikes}
We consider the scenario of IMBHs originated  from massive objects formed directly during
the collapse of primordial gas (pGas) in early-forming halos.
The initial DM profile of the halo prior to  the adiabatic growth of the blackhole  is 
assumed to be NFW [20] 
\begin{align}
\rho(r)=\frac{\rho_{s}}{\left( \frac{r}{r_{s}} \right)^{\gamma} \left( 1+\frac{r}{r_{s}} \right)^{3-\gamma} }  ,
\end{align}
where $\rho_{s}$ and $r_{s}$ are reference energy density and distance, respectively.
$\gamma$ is the inner slope of the profile, and in the standard NFW profile $\gamma=1$.
After the adiabatic growth of the blackhole, the spiked profile is a power law%
[21] 
\begin{align}
\rho_{sp}(r)=\rho(r_{sp})\left( \frac{r}{r_{sp}}\right)^{-\gamma_{sp}} ,
\end{align}
where
$r_{sp}\sim \tx{pc}$ and
$\gamma_{sp}=(9-2\gamma)/(4-\gamma)$.
For $\gamma=1$, $\gamma_{sp}=7/3$.
The spiked DM profile is apparently divergent at small radii.
However, the DM annihilation sets an upper limit $\rho_{\text{lim}}$ on the DM density
of order 
$\rho_{\text{lim}}=\rho_{sp}(r_{\text{lim}})\approx m_{\chi}/\langle\sigma v\rangle t$ with $t$ the age of the mini-spike.
Thus $\rho_{sp}(r)$ should have an inner cut-off at radius $r_{\text{cut}}$,
which is defined as
$r_{\text{cut}}=\text{max}\{4 R_{\text{Schw}}, r_{\text{lim}}\}$,
where $R_{\text{Schw}}$ is the Schwarzschild radius of the IMBH.
For typical value of $r_{\text{lim}}\approx 10^{-3}$~pc, 
which is larger than the typical Schwartzchild radius  
$R_{\text{Schw}}\approx 2.95~\text{km}~ (M_{bh}/M_{\odot})$.
Thus we take $r_{\tx{cut}}=r_{\tx{lim}}$.
The source term from the annihilation of Majorana DM particles (with mass $m_{\chi}$) 
directly into $e^{+}e^{-}$ can be written as
\begin{align}
Q_{e}(E)=\frac{N_{e} \langle\sigma v\rangle}{2 m_{\chi}^{2}}  
\delta(E-E_{0})
\int_{r_{\text{cut}}}^{r_{sp}} \rho_{sp}^{2}(r') 4\pi r'^{2} dr' ,
\end{align}
where $N_{e}=2$ is the number of CRE per DM annihilation, and  
$E_{0}=m_{\chi}$.
After integrating over $r'$, the source term can be written as 
\begin{align}
Q_{e}(E)=\frac{N_{e} \langle\sigma v\rangle L}{2 m_{\chi}^{2}} 
\delta(E-E_{0})  ,
\end{align}
where  $L$ is the annihilation luminosity
\begin{align}
L
\equiv\int_{r_{\text{cut}}}^{r_{sp}} \rho_{sp}^{2}(r') 4\pi r'^{2} dr' 
=\frac{4\pi r_{sp}^{3}}{2\gamma_{sp}-3}\rho^{2}(r_{sp}) 
\left( \frac{r_{\text{cut}}}{r_{sp}}\right)^{3-2\gamma_{sp}} .
\end{align}
In the case of  $\gamma_{sp}=7/3$, it can be written as
\begin{align}
Q_{e}(E)=\frac{6 \pi N_{e} \langle\sigma v\rangle}{5 m_{\chi}^{2}} 
\rho^{2}(r_{sp}) r_{sp}^{14/3} r_{\text{cut}}^{-5/3} 
\delta(E-E_{0})  ,
\end{align}
which leads to the expression  of Eq.~(3) 
in the letter.

\subsubsection{DM subhalos}
The DM subhalos located at  the inner volume of the Galaxy
may experience a significant degree of mass loss due to tidal stripping,
as they encounter other subhalos more frequently.
Therefore we  adopt a tidally truncated  density profile for DM subhalos
[25-27].
\begin{align}
\rho(r)	
=
\frac{\rho_{0}}{(r/\tx{kpc})^{\gamma}}e^{-r/R_{b}}  .
\end{align}
The parameters  $\rho_{0}$, $\gamma$ and $R_{b}$ can be parametrized as 
$\gamma(d)=g_{1}(d/\text{kpc})^{g_{2}}$ and $R_{b}=b_{1}(d/\text{kpc})^{b_{2}}$,
where $d\approx 8.5$~kpc is  
the distance from the center of the subhalo to the GC.
The values of the parameters $g_{1,2}$ and $b_{1,2}$   
can be extrapolated from the N-body simulations
for different ranges of the total mass $M_{h}$. 
For fixed $r$ and $R_{b}$,
the normalization factor $\rho_{0}$ can be  determined by the subhalo total mass $M_{h}$.
Based on  a joint analysis to the Via Lactea II and  ELVIS simulations
in Ref. [27],
we obtain the values of $\gamma$, $R_{b}$ and $\rho_{0}$ for selected values of 
$M_{h}$, which are listed in \tab{tab:subhalo-profile}.

\begin{figure}[htb]
	\centering
	\includegraphics[width=0.4\textwidth]{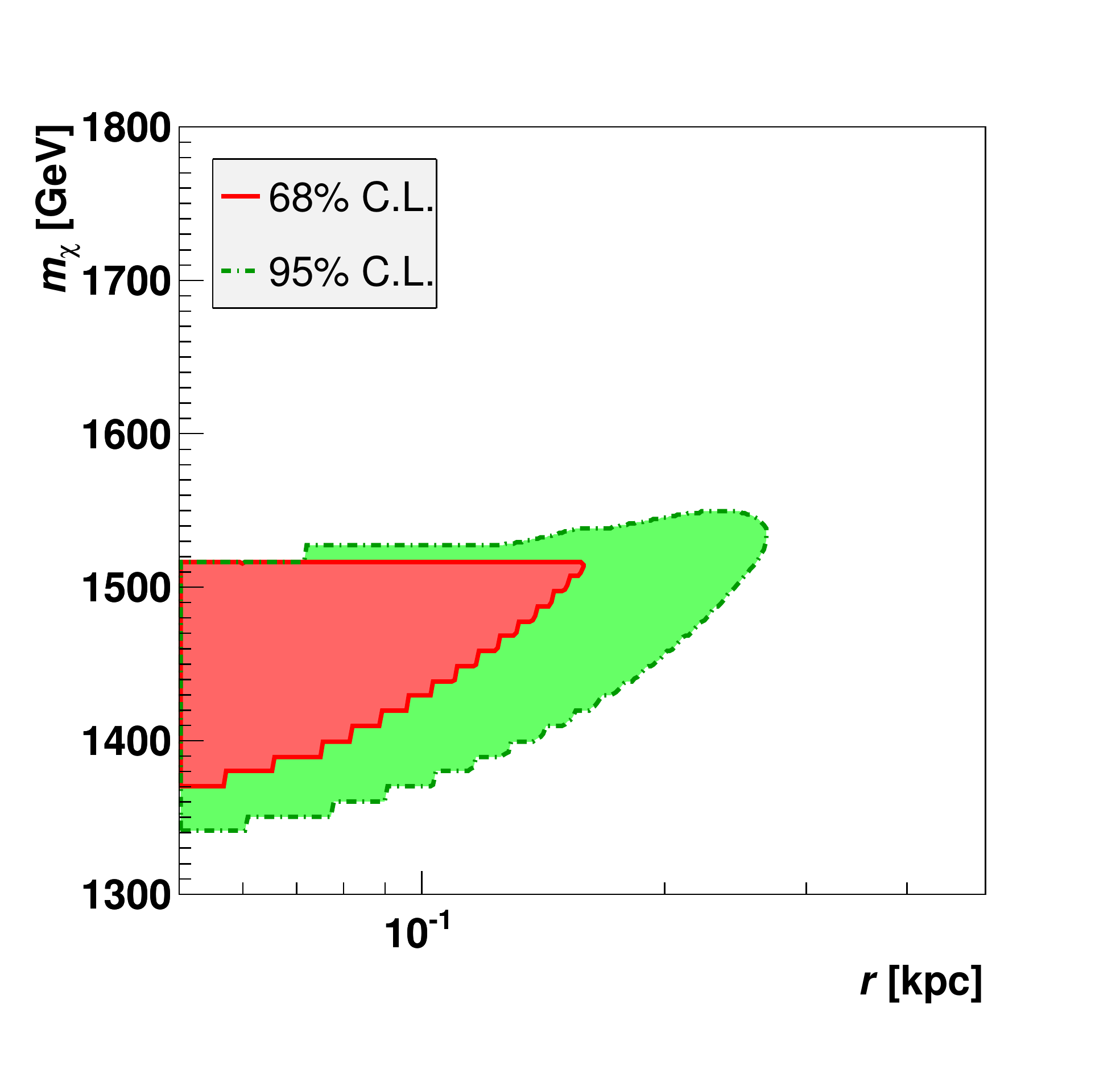}
	\includegraphics[width=0.4\columnwidth]{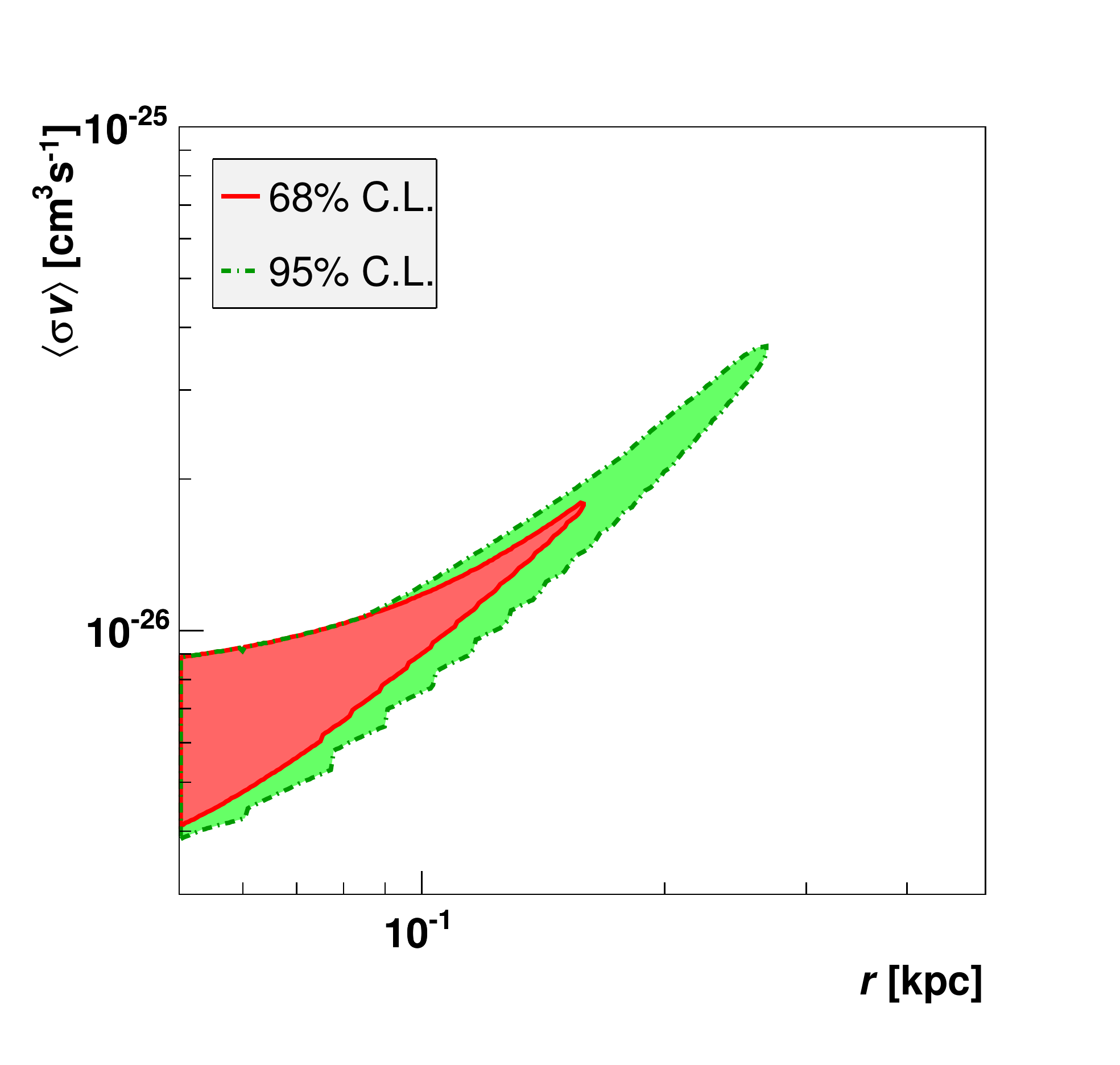}
	\caption{		
	Left) 
	Allowed regions at $68\%$ and $95\%$ C.Ls. in $(r, m_{\chi})$ plane for 
	the source of DM subhalos.
	Right)
	Allowed region in $(r, \langle \sigma v\rangle)$ plane.	
	}\label{fig:extendDM-contour}
\end{figure}

\begin{table}[htb]
	\centering
		\begin{tabular}{c|ccc}
		\hline\hline
		$M_{h}~(M_{\odot})$ & $\gamma$ & $R_{b}~\text{(kpc)}$ & $\rho_{0}$\\
		\hline
		$1\times 10^{6}$ 	& 0.81 	&	0.054	& 1.8\\
		$1\times 10^{7}$ 	& 0.78 	&	0.096 	& 5.3 \\
		$3\times 10^{7}$ 	& 0.65	&	0.19         & 4.5 \\
		$1\times 10^{8}$ 	& 0.75	&	0.22         & 9.4 \\
		\hline\hline
		\end{tabular}
	\caption{Prameters of DM subhalo density profiles extrapolated to the distance $d\approx 8.5$~kpc 
	for difference choices of the subhalo masses
	from [27]. 
	The reference density $\rho_{0}$ is in units of $\tx{GeV}\cdot\tx{cm}^{-3}$.
	}
\label{tab:subhalo-profile}
\end{table}
For a typical subhalo masses $M_{h}=10^{7} M_{\odot}$,  
the best-fit parameters 
for three choices of ($r$,$m_{\chi}$) are shown in \tab{tab:extendDM},
which correspond to the three cases plotted in the middle panel of Fig.~2. 
Similar to the case of continuous point-like sources,
when $r$ is allowed to vary freely, 
the value of $\chi^{2}$ decreases with decreasing distance and
approaches a minimal value $\chi^{2}_{\text{min}}=13.2$  in the limit of  $r\to 0$. 
The  allowed regions at $68\%$ ($95\%$) C.L. in $(r, m_{\chi})$ and $(r, \sigmav)$ planes 
are shown  in \fig{fig:extendDM-contour}.
The allowed range of distance is $r \lesssim 0.3$~kpc at $95\%$~C.L., 
which is almost the same as that in the case of mini-spikes.
The fit results for other subhalo masses 
$M_{h}=10^{6} M_{\odot}$, $3\times 10^{7} M_{\odot}$ and $10^{8} M_{\odot}$
are shown separately in \tab{tab:extendDM2}.
The results show that the DAMPE excess can be well fitted for a large range of $M_{h}$.
Increasing the value of $M_{h}$ leads to a decrease of the DM annihilation cross section.

\begin{table}[htbp]
	\centering
		\begin{tabular}{cc|cccc|c}
		\hline\hline
		  $r$ [kpc] & $m_\chi$[GeV] & $\langle \sigma v \rangle$  & $\gamma_1$ & $\gamma_2$ & $N_b$  & $\chi^2/\tx{d.o.f.}$ \\
				 \hline
				 0.1& 1510 & $ 1.04\pm0.30$  & $3.10\pm0.01$ & $4.08^{+0.14}_{-0.13}$ & $1.90\pm0.03$ & 13.37/28\\
		0.2 &1520 & $2.55^{+0.82}_{-0.83}$ & $3.10\pm0.01$& $4.10^{+0.15}_{-0.13}$&  $1.90\pm0.03$ & 15.90/28\\
		0.3 &1550 & $4.62^{+1.83}_{-1.87}$ & $3.10\pm0.01$& $4.10^{+0.16}_{-0.14}$& $1.89\pm0.03$& 19.20/28\\
		  \hline\hline
		\end{tabular}
	\caption{
	Best-fit parameters for three choices of $(r, m_{\chi})$ values for a source of  DM subhalo 
	with mass $M_{h}=10^{7} M_{\odot}$. The corresponding subhalo density parameters are listed
	in \tab{tab:subhalo-profile}.  
	$\langle \sigma v\rangle$ is in units of $10^{-26}\text{cm}^{3}\text{s}^{-1}$ and 
	$N_{b}$ is in units of $10^{-7}$~GeV$^{-1}$m$^{-2}$s$^{-1}$sr$^{-1}$.
	}
\label{tab:extendDM}
\end{table}

\begin{table}[htbp]
	\centering
		\begin{tabular}{ccc|cccc|c}
		\hline\hline
		 $M [M_\odot]$ & $r$ [kpc] & $m_\chi$[GeV] & $\langle \sigma v \rangle$  & $\gamma_1$ & $\gamma_2$ & $N_b$  & $\chi^2/\tx{d.o.f.}$ \\
		  \hline
				$10^6$ & 0.1& 1510 & $ 15.15\pm4.32$  & $3.10\pm0.01$ & $4.08^{+0.14}_{-0.13}$ & $1.90\pm0.03$ & 13.26/28\\
		-- & 0.2 &1520 & $39.88^{+12.79}_{-12.83}$ & $3.10\pm0.01$& $4.10^{+0.15}_{-0.13}$&  $1.90\pm0.03$ & 15.81/28\\
		-- & 0.3 &1550 & $72.9^{+28.73}_{-29.29}$ & $3.10\pm0.01$& $4.10^{+0.16}_{-0.13}$& $1.89\pm0.03$& 19.15/28\\ 
		\hline
		$3\times10^7$ & 0.1& 1510 & $ 1.49\pm0.44$  & $3.10\pm0.01$ & $4.08^{+0.14}_{-0.13}$ & $1.90\pm0.03$ & 13.88/28\\
		-- & 0.2 &1520 & $2.95^{+0.93}_{-0.94}$ & $3.10\pm0.01$& $4.11^{+0.15}_{-0.14}$&  $1.90\pm0.03$ & 16.43/28\\
		-- & 0.3 &1550 & $4.93^{+2.03}_{-2.05}$ & $3.10\pm0.01$& $4.11^{+0.16}_{-0.14}$& $1.89^{+0.03}_{-0.04}$& 19.60/28\\
		\hline
		$10^8$ & 0.1& 1510 & $ 0.17\pm0.05$  & $3.10\pm0.01$ & $4.08^{+0.14}_{-0.13}$ & $1.90\pm0.03$ & 13.89/28\\
		-- & 0.2 &1520 & $ 0.33\pm0.11$ & $3.10\pm0.01$& $4.11^{+0.15}_{-0.14}$&  $1.90\pm0.03$ & 16.46/28\\
		-- & 0.3 &1550 & $0.55^{+0.22}_{-0.23}$ & $3.10\pm0.01$& $4.11^{+0.16}_{-0.14}$& $1.89^{+0.03}_{-0.04}$& 19.64/28\\
		\hline\hline
		\end{tabular}
	\caption{
	The same as \tab{tab:extendDM}, but with different DM subhalo masses
	$M_{h}=10^{6} M_{\odot},~3\times 10^{7}M_{\odot}$ and $10^{8} M_{\odot}$.
	The corresponding DM subhalo density parameters are listed
	in \tab{tab:subhalo-profile}.
	 }
\label{tab:extendDM2}
\end{table}

\subsubsection{Contributions from the Galactic halo DM}
\begin{figure}[!htb]
	\centering
	\includegraphics[width=0.8\columnwidth]{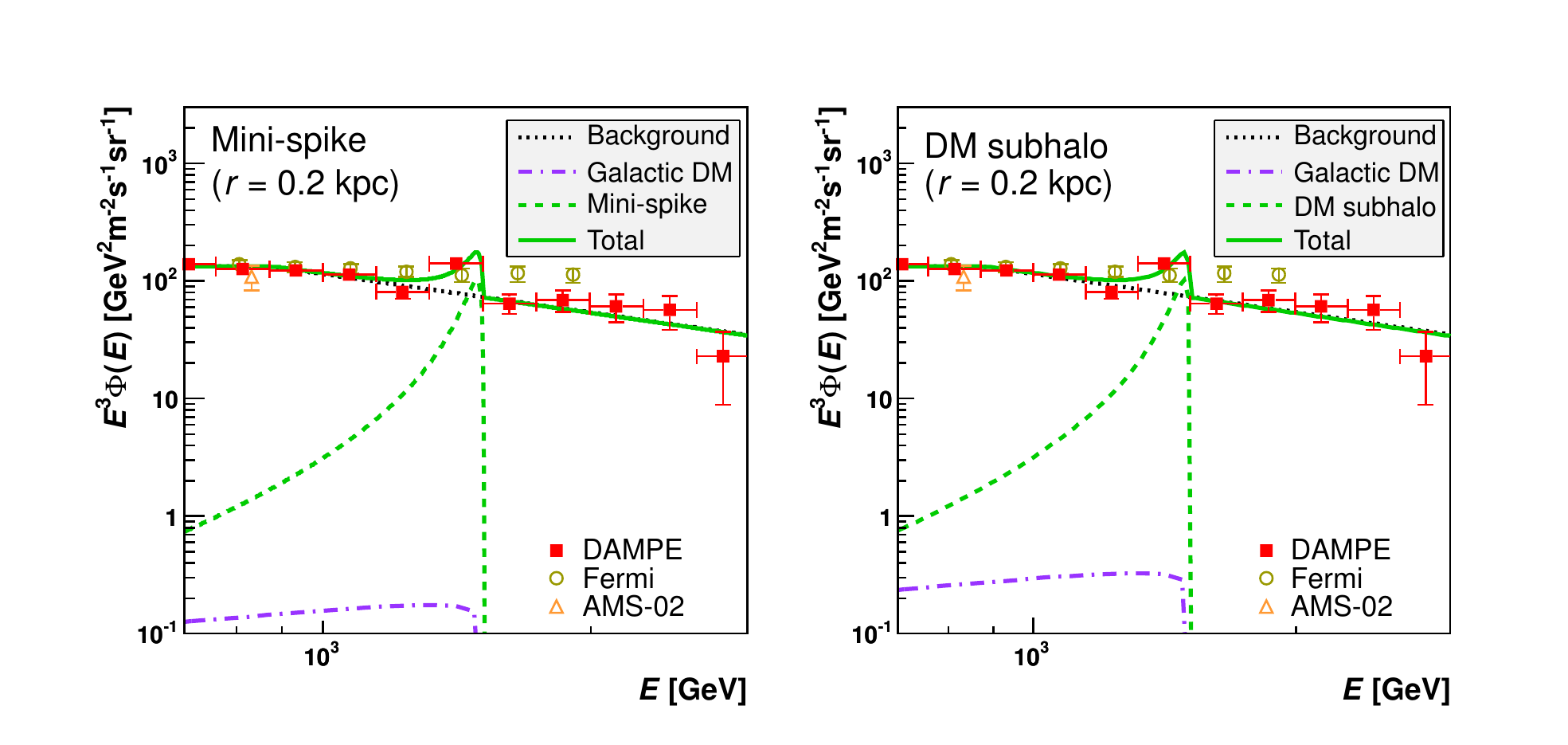}
	\caption{
	Left)
	Comparison of the CRE energy spectrum from a mini-spike and that from the whole Galactic halo DM
	with the same DM annihilation cross section. 
	The parameters of the mini-spike correspond to the case of $r=0.2$~kpc in \tab{tab:pointDM}.
	Right)
	Comparison of  electron energy spectrum from a nearby DM subhalos 
	and that from the whole Galactic halo DM with the same annihilation cross section. 
	The parameters of the DM subhalo correspond to the case of $r=0.2$~kpc in \tab{tab:extendDM}.
	}
	\label{fig:DM-Galactic-halo}
\end{figure}

For DM sources, the contributions to CRE also come from the whole Galactic DM halo.
If the favoured  DM annihilation cross sections are around the thermal value
$\sigmav_{F}=3\times 10^{-26}~\tx{cm}^{3}\tx{s}^{-1}$,
the contributions to the CRE flux  from the whole Galactic DM above TeV
are found  to be  negligibly small. 
In \fig{fig:DM-Galactic-halo}, we compare the contributions
to CRE from DM substructures and DM Galactic halo for typical cases.
For calculating the contributions from Galactic DM halo,
we make use of the GALPROP v54 package [32] 
with a reference propagation parameter set
$^SL^Z6^R20^T100000^C5$ [33] %
and the Einasto DM profile with a local DM density of 
$0.4~\text{GeV}\text{cm}^{-3}$.
The contributions from remote DM subhalos in the Galaxy can be safely  neglected, 
as the N-body simulations show that they contribute only to a tiny  fraction of 
the total Galactic DM mass [22-24]. %

\subsection{C. Burst-like sources}
\begin{figure}[htb]
	\centering
	\includegraphics[width=0.4\columnwidth]{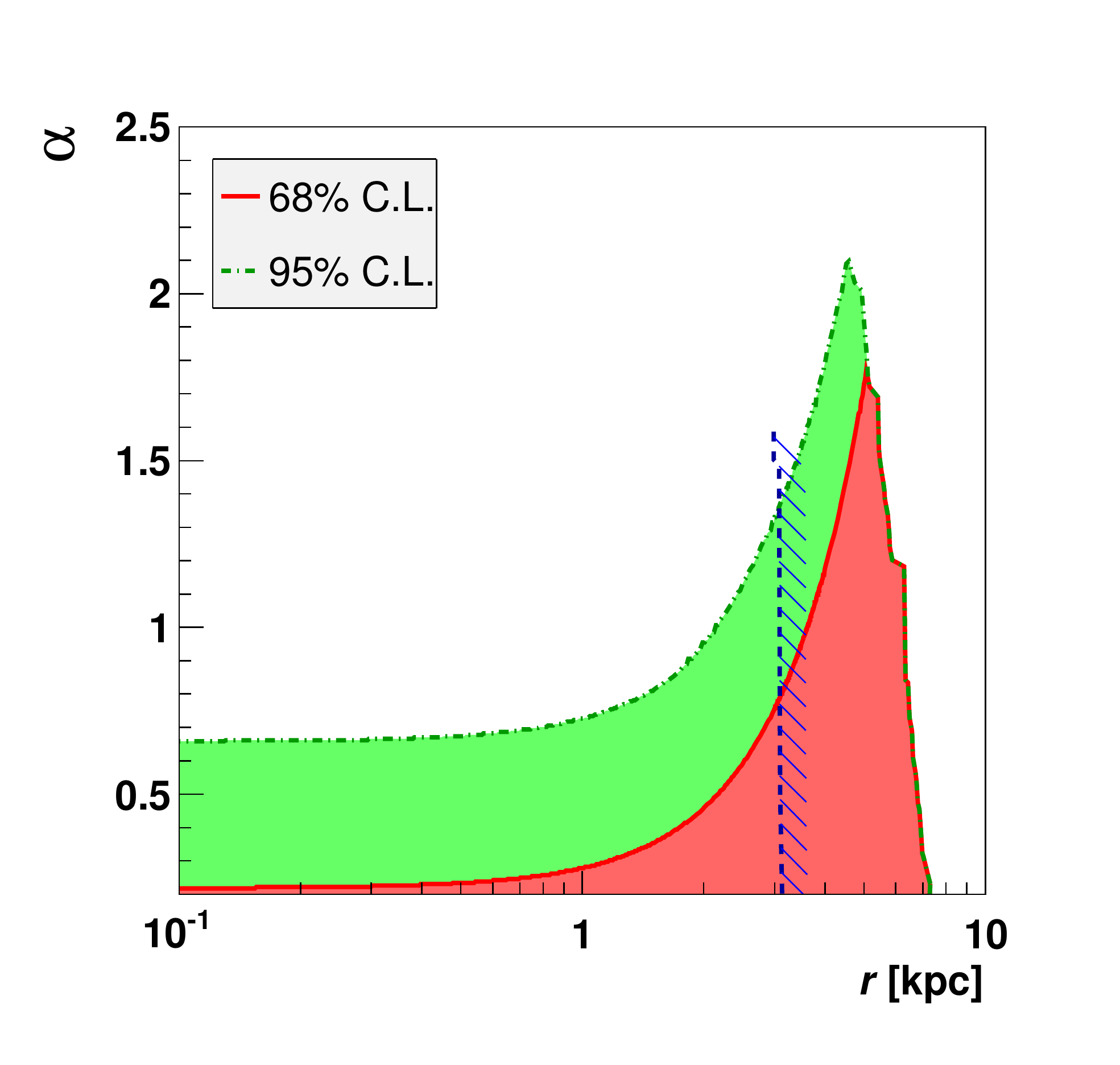}	
	\caption{
	 Allowed regions at $68\%$ and $95\%$ C.L. in $(r, \alpha)$ plane for the burst-like point source.
	 The region excluded at large distances    $r\gtrsim 3$~kpc is due to the requirement that 
	 the total energy $E_{\text{tot}} <10^{51}$~erg.
	}\label{fig:r-alpha-dependenc}
\end{figure}

For the burst-like sources, 
the best-fit parameters for three representative choices  of 
$(r, \alpha)$ shown in the right panel of Fig.~2 in the letter  
are summarized  in  \tab{tab:SNR},
where the total released energy $E_{\text{tot}}$ is derived from the value of $N_{0}$,
using
$E_{\text{tot}}=\int_{E_{\tx{min}}} E Q(r,t,E) dE d^{3}r dt$ with a lower cutoff $E_{\tx{min}}=0.1$~GeV.
When $r$ and $\alpha$ are allowed to vary freely in the fit, 
the $\chi^{2}$ gradually approaches a minimal value  of  12.5 for $\alpha\to 0$.
The allowed regions in ($r$, $\alpha$) at $68\%$ and $95\%$~C.L.
are shown in  \fig{fig:r-alpha-dependenc}.
In the region $r \leq r_{d}(E_{\text{max}})$, 
the favoured $\alpha$ is less than 0.65 at $95\%$~C.L. which is highly 
insensitive to the distance, as the spectral shape is dominated by the effect
of ``phase-space shrinking''. 
In the region $r > r_{d}(E_{\text{max}})$, the effect of energy-dependent diffusion
becomes important and the allowed value of $\alpha$ can be larger which can reach 
1.3 (1.6) at $r\approx 3 (4)$~kpc.
By imposing the condition of $E_{\text{tot}}<10^{51}(10^{53})$~erg,
the distance $r$ is restricted in range $r\lesssim 3$(4)~kpc.
Together with the required  $t=(1.5-1.7)\times 10^{5}$~yr,
we find the 7 candidate pulsars in the ATNF catalogue.
The relevant parameters of these candidates are summarized in \tab{tab:ATNF}.

\begin{table}[htbp]
	\centering
		\begin{tabular}{cc|ccccc|c}
		\hline\hline
$r$ [kpc] & $\alpha $& $t$ (Myr) & $N_0\ (\text{GeV}^{-1})$ & $\gamma_1$ & $\gamma_2$ & $N_b$  & $\chi^2/\text{d.o.f.}$ \\
		  \hline
		  1.0 & 0.5 & $0.147^{+0.002}_{-0.003}$ & $ 3.72^{+1.18}_{-1.21}\times10^{44}$  & $3.10\pm0.01$ & $4.09^{+0.15}_{-0.13}$ & $1.90\pm0.03$ & 15.56/27\\
		  2.0 & 0.7 & $0.147^{+0.002}_{-0.003}$  & $2.33^{+0.74}_{-0.79}\times10^{46}$  & $3.10\pm0.01$ & $4.09^{+0.15}_{-0.13}$ & $1.90\pm0.03$ & 15.62/27\\
		  3.0 & 1.3 & $0.146^{+0.002}_{-0.003}$  & $3.10^{+1.13}_{-1.26}\times10^{50}$  &  $3.10\pm0.01$ & $4.10^{+0.16}_{-0.14}$ & $1.90\pm0.03$ & 17.51/27\\
		\hline\hline
		\end{tabular}
	\caption{ 
	Best-fit parameters for three representative choices of $(r, \alpha)$ in the case of 
	burst-like sources. The cutoff energy is fixed at $E_c=5\times10^{4}$ GeV.
	$N_b$ is in units of $10^{-7}$GeV$^{-1}$m$^{-2}$s$^{-1}$sr$^{-1}$. 
	The total released energy for the cases of $r=1$, 2 and 3~kpc is 	
	 $ E_{tot}=2.53^{+0.80}_{-0.82}\times10^{48}$,
	 $2.18^{+0.70}_{-0.74}\times10^{49}$ and
	 $9.56^{+3.49}_{-3.88}\times10^{50}$~erg, respectively.
	}
\label{tab:SNR}
\end{table}

\begin{table}[htbp]
	\centering
		\begin{tabular}{l|cc|cccc} 
		\hline\hline
		  Source Name & $\ell$(deg) &  $b$(deg) & $r$ (kpc) & $t$ (yr) & $\dot{E}$ (erg/s)  \\
		\hline
		B0740-28 & 243.77 & -2.44 & 2.00  & $1.57\times10^{5}$ & $1.4\times10^{35}$  \\
		J0922-4949 & 272.24 & 0.16 &  2.70  & $1.54\times10^{5}$ & $4.5\times10^{33}$  \\
		J1055-6022 & 289.11 & -0.65 & 3.60 & $1.62\times10^{5}$ & $4.3\times10^{33}$  \\
		J1151-6108 & 295.81 & 0.91  & 2.22  & $1.57\times10^{5}$ & $3.9\times10^{35}$  \\
		J1509-5850 & 319.97 & -0.62 & 3.35 & $1.54\times10^{5}$ & $5.1\times10^{35}$  \\
		J1616-5017 & 332.83 & 0.29 & 3.48  & $1.67\times10^{5}$ & $1.6\times10^{34}$ \\
		J1739-3023 & 358.09 & 0.34 & 3.07 & $1.59\times10^{5}$ & $3.0\times10^{35}$ \\
				\hline\hline
	 \end{tabular}%
	 \caption{
	 	Parameters 
		of candidate pulsars  in ATNF pulsar catalog [30] 
		satisfying the conditions of 
		 age $t=0.15-0.17$~Myr  as required to reproduce 
		 the DAMPE excess, and distance $r<4$~kpc from the requirement of $E_{\tx{tot}}<10^{53}$~erg.
		The quantities  $(\ell,b)$, $r$, $t$ and $\dot{E}$ stand for the 
		 direction, distance, age and spin-down energy loss rate, respectively.
	 }
	\label{tab:ATNF}%
\end{table}%

\newpage
\subsection{D. Electron anisotropies}
\begin{figure}[htb]
\begin{center}
\includegraphics[width=1\textwidth]{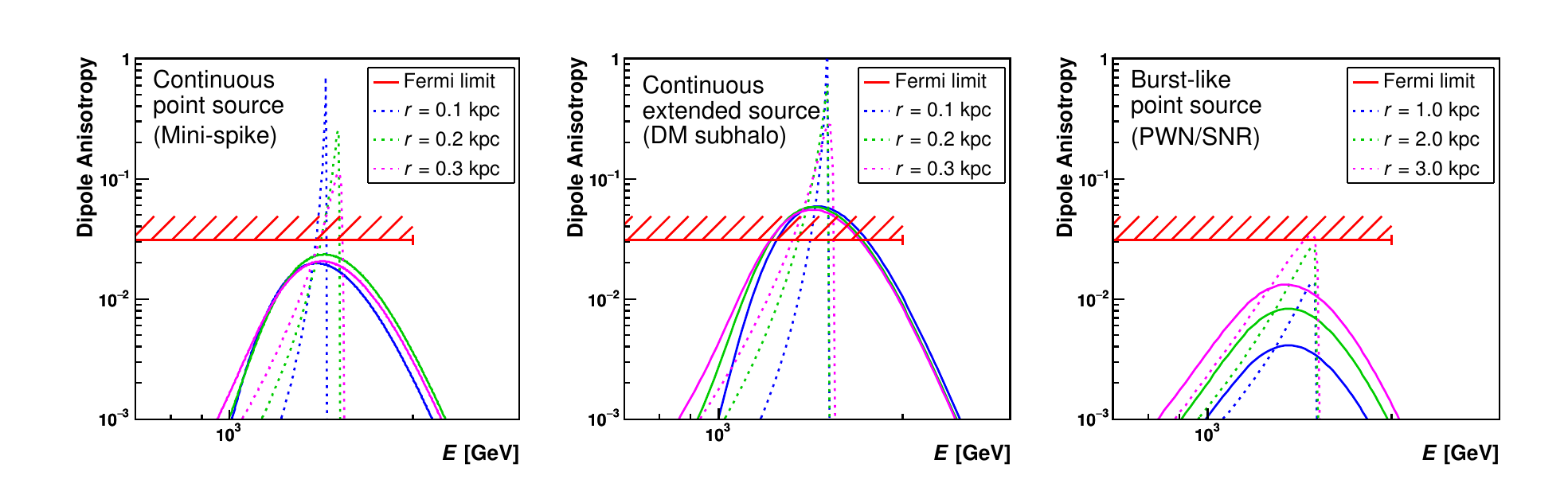}
\caption{
Predictions for electron anisotropies in the arrival directions  (dotted curves).
Left) for continuous point-like sources with three parameter sets 
listed in \tab{tab:pointDM}
Middle) for continuous extended sources with parameter sets 
listed in \tab{tab:extendDM}.
Right) for burst-like sources with  parameter sets listed in \tab{tab:SNR}.
For all the type of sources, 
the dashed curves correspond to the anisotropies convoluted with an energy resolution of $15\%$,
and the horizontal lines indicate the averaged value in the energy bin  0.55-2~TeV, 
the corresponding  current upper limits from Fermi-LAT are also shown.
}
\label{fig:anisotropy_DAMPE}
\end{center}
\end{figure}

For one or a few nearby sources,
the anisotropy in the arrival directions of CREs 
is dominated by the dipole term which can be approximated as 
\begin{align}
\Delta \approx \frac{3 D(E)}{c}
\left|\frac{\bigtriangledown \Phi_{\text{tot}}}{\Phi_{\text{tot}}}\right|  ,
\end{align}
where  
$\Phi_{\text{tot}}$ is the total flux of the nearby sources and the background,
and $c$ is the speed of light.
In the approximation that the background is nearly isotropic, %
the anisotropy  for the continuous point-like sources  is given by 
\begin{align}
\Delta_{\text{conti}} \approx\frac{3r}{2c}\frac{(1-\delta)b_{0}E }{[1-(E/E_{0})^{1-\delta}]}
\frac{f(r,E)}{f_{\text{tot}}(r,E)}  ,
\end{align}
where $f_{\text{tot}}=f(r,E)+f_{bg}(r,E)$ is the total distribution function.
For the burst-like source it can be written as
\begin{align}
\Delta_{\text{burst}} \approx\frac{3 r}{2ct}\frac{(1-\delta) E/E_{\text{max}}}{1-\xi(E)^{1-\delta}}
\frac{f(r,E)}{f_{\text{tot}}(r,E)}  .
\end{align}
The predicted anisotropies according to the best-fit parameters in each case of 
the three type of sources listed in \tab{tab:pointDM}, \tab{tab:extendDM} and \tab{tab:SNR} are shown 
in \fig{fig:anisotropy_DAMPE}.
The anisotropy of the background CRE depends on the assumed spatial distribution of 
primary CRE sources, the typical values estimated from GALPROP is of $\mathcal{O}(10^{-3})$
[31]
which is significantly smaller than that generated from nearby sources.

\subsection{E. Associated gamma-ray signals}
\subsubsection{Prompt gamma-rays}
Prompt photons are generated from DM annihilation through FSR of final 
state charged leptons, which is of particular importance for electron final states.
The flux of FSR photon from the annihilation of Majorana DM particles (with mass $m_{\chi}$)
per solid angle $d\Omega$ is given by
\begin{align}
\frac{d\Phi_{\gamma}}{d\Omega dE_{\gamma}}
=
\frac{\langle \sigma v \rangle}{8\pi m_{\chi}^{2}} 
\frac{dN_{\gamma}}{dE_{\gamma}}
\int_{\text{l.o.s}} \rho^{2}(r') ds  .
\end{align}
where $\rho(r')$ is the DM density profile,
and the integration is performed along the light-of-sight $s$ which is related to $r'$ as
$r'=\sqrt{s^{2}+r^{2}-2 s r \cos\theta}$ with $r$ the distance from the detector to the center of the DM halo
and $\theta$ is the angle away from the center.
The spectrum  for $e^{+}e^{-}$ final state is given by
\begin{align}
\frac{dN_{\gamma}}{d x}
=
\frac{\alpha_{\text{em}}}{\pi}
\frac{1+(1-x)^{2}}{x}
\left[ 
	-1+
	\ln\left( \frac{4(1-x)}{\epsilon^{2}}\right)
\right]  ,
\end{align} 
where  $x=E_{\gamma}/m_{\chi}$, $\epsilon=m_{e}/m_{\chi}$ and
$\alpha_{\text{em}}$ is the fine structure constant.
An important feature of the FSR photon spectrum is that 
it can be approximated as a power law at low energies
\begin{align}
\frac{dN_{\gamma}}{d E}\propto E^{-1}, 
\ \ \ \ \ (~\tx{for} E_{\gamma}\ll m_{\chi}~)
\end{align}
and has a sharp cutoff at $E_{\gamma}=m_{\chi}$.
In numerical calculations we use the package PYTHIA 8.1 which 
adopts the same formula.
Since mini-spikes are point-like, 
the flux of prompt $\gamma$-rays from DM annihilation in mini-spikes 
can be approximated as
\begin{align}
\frac{d\Phi_{\gamma}}{dE_{\gamma}}=
\frac{\langle\sigma v\rangle}{2 m_{\chi}^{2}} \frac{L}{4\pi r^{2}}
\frac{dN_{\gamma}}{dE_{\gamma}} ,
\end{align}
where $L$ is the annihilation luminosity given in the letter,
and $r$ is the distance from the detector to the center of the mini-spike.
The prompt $\gamma$-rays from DM subhalos  and Galactic DM are calculated by
direct integration over the DM density profile.
Unlike the case of the CRE, for prompt $\gamma$-rays, the contribution from the 
whole Galactic halo  DM is significant.

\subsubsection{ICS  gamma-rays}

The distribution of final state photon from the  ICS process 
$e(E_{e})+\gamma(E_{\gamma})\rightarrow e'(E'_{e})+\gamma'(E'_{\gamma})$, 
where $E_{e}(E'_{e})$, $E_{\gamma}(E'_{\gamma})$ are the energies of initial (final) 
state electron and photon, respectively,  is given by 
\begin{equation}
\frac{dN_{\gamma'}}{dE_{\gamma'}dt}=2\pi \alpha^{2}_{\text{em}}\frac{u_{\gamma}}{E_e^2 E_{\gamma}^2}f_{\tx{IC}}(q,\varepsilon), 
\end{equation}
where $u_{\gamma}$ is the energy density of the initial photons.
The function $f_{\tx{IC}}(q,\epsilon)$ is 
\begin{equation}
f_{\tx{IC}}(q,\varepsilon)=2q\ln(q)+(1+2q)(1-q)+\frac{(\varepsilon q)^2}{2(1+\varepsilon q)}(1-q)  ,
\end{equation}
where 
$\varepsilon=E_{\gamma}'/E_e$, 
$\Gamma=4E_{\gamma}E_e/m_e^2$  and  
$q=\varepsilon/\Gamma(1-\varepsilon)$.
The final energy  satisfies the relation
$E_{\gamma}/E_e\leq \varepsilon \leq \Gamma/(1+\Gamma)$.
The photon flux $\Phi_{\gamma}=dN_{\gamma}/dA dt$ obtained for 
a given line of sight is [42] %
\begin{equation}
\frac{d^2\Phi_{\gamma'}}{dE_{\gamma'}d\Omega}
=
\frac{1}{2}\alpha_{\text{em}}^2  \int_{\text{l.o.s}}\,ds \int\int \frac{dE_e}{E_e^2}\frac{dE_{\gamma}}{E_{\gamma}^2}
f(r,E_{e})%
u_{\gamma}(E_{\gamma})
f_{\tx{IC}}  ,
\end{equation}
where
$f(r,E_{e})$ is the density of initial state electrons.
The energy density $u_{\gamma}(E_{\gamma})$ is assumed to be black body like
\begin{equation}
u_{\gamma}(E_{\gamma})=\mathcal{N}\frac{E_{\gamma}^3}{\pi^2\left(e^{E_{\gamma}/T}-1\right)} ,
\end{equation}
where $\mathcal{N}$ is a position-dependent normalization constant.
The three major components of the ISRF are
cosmic microwave background with temperature
$T_{\text{CMB}}=2.35\times 10^{-4}$~eV,
the infrared radiation produced by the absorption and  
re-emission of star light by 
the interstellar dust
with temperature
$T_{\text{dust}}=3.5\times 10^{-3}$~eV
and the start light with temperature
$T_{\text{star}}=0.3$~eV.
For the nomalization factors, we use the values
$\mathcal{N}_{\text{CMB}}=1$, 
$\mathcal{N}_{\text{dust}}=2.5\times 10^{-5}$ and 
$\mathcal{N}_{\text{star}}=6.0\times 10^{-13}$, respectively,
which are the values at the solar neighbourhood $d\approx 8$~kpc interpolated from [34].

\subsubsection{Gamma-ray signals of  mini-spikes, DM subhalos and burst-like sources}
\begin{figure}[!htbp]
\begin{center}
\includegraphics[width=0.8\textwidth]{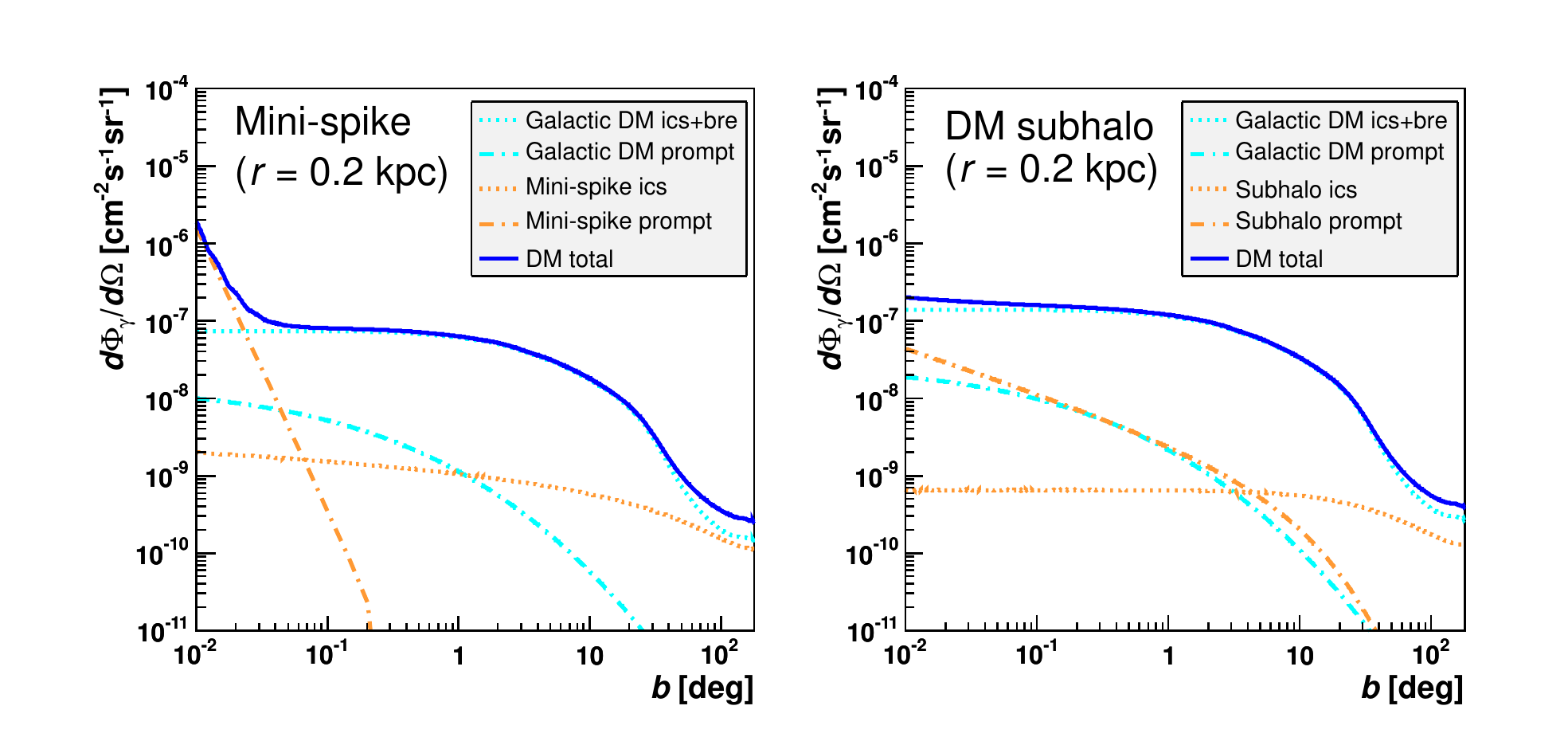}
\includegraphics[width=0.8\textwidth]{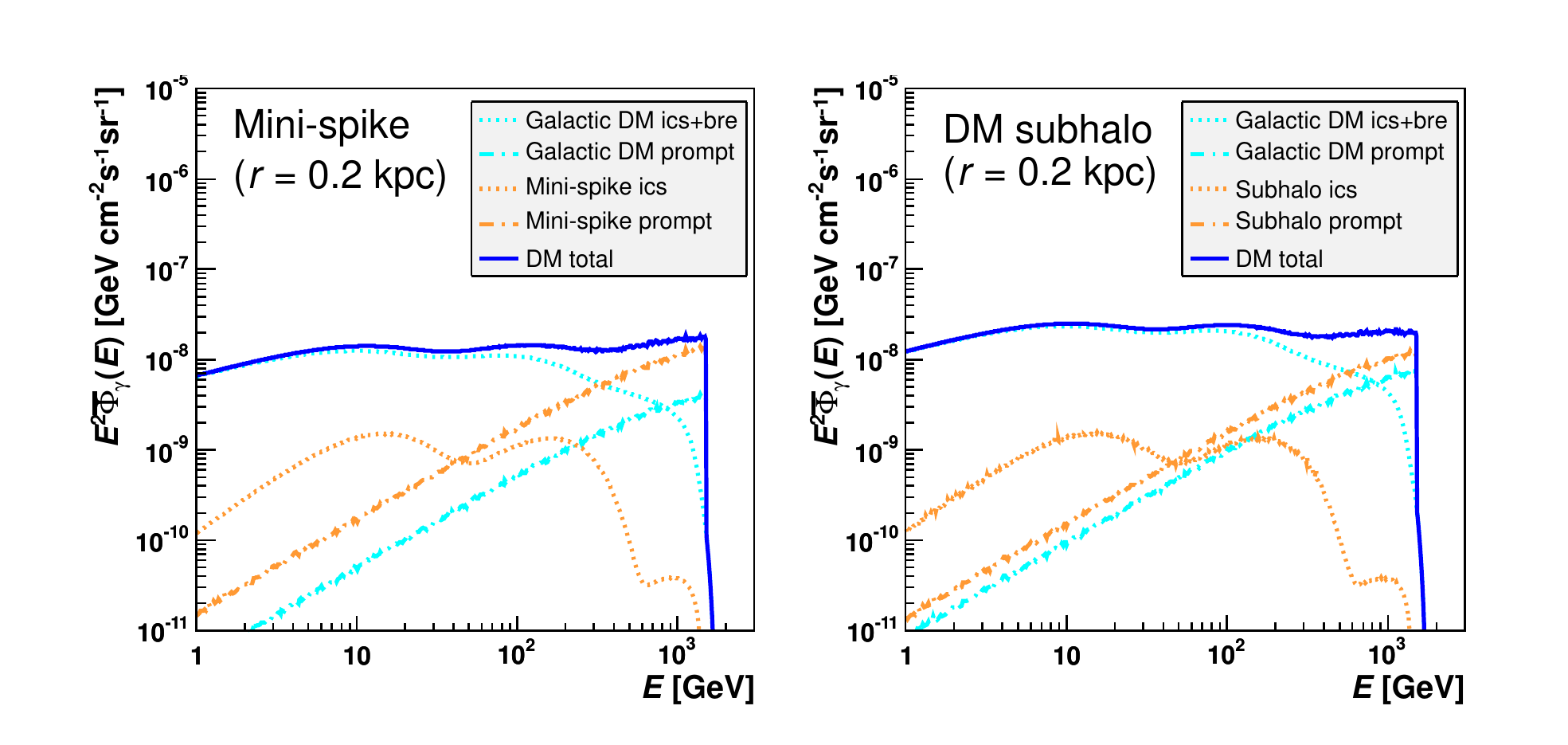}
\includegraphics[width=0.8\textwidth]{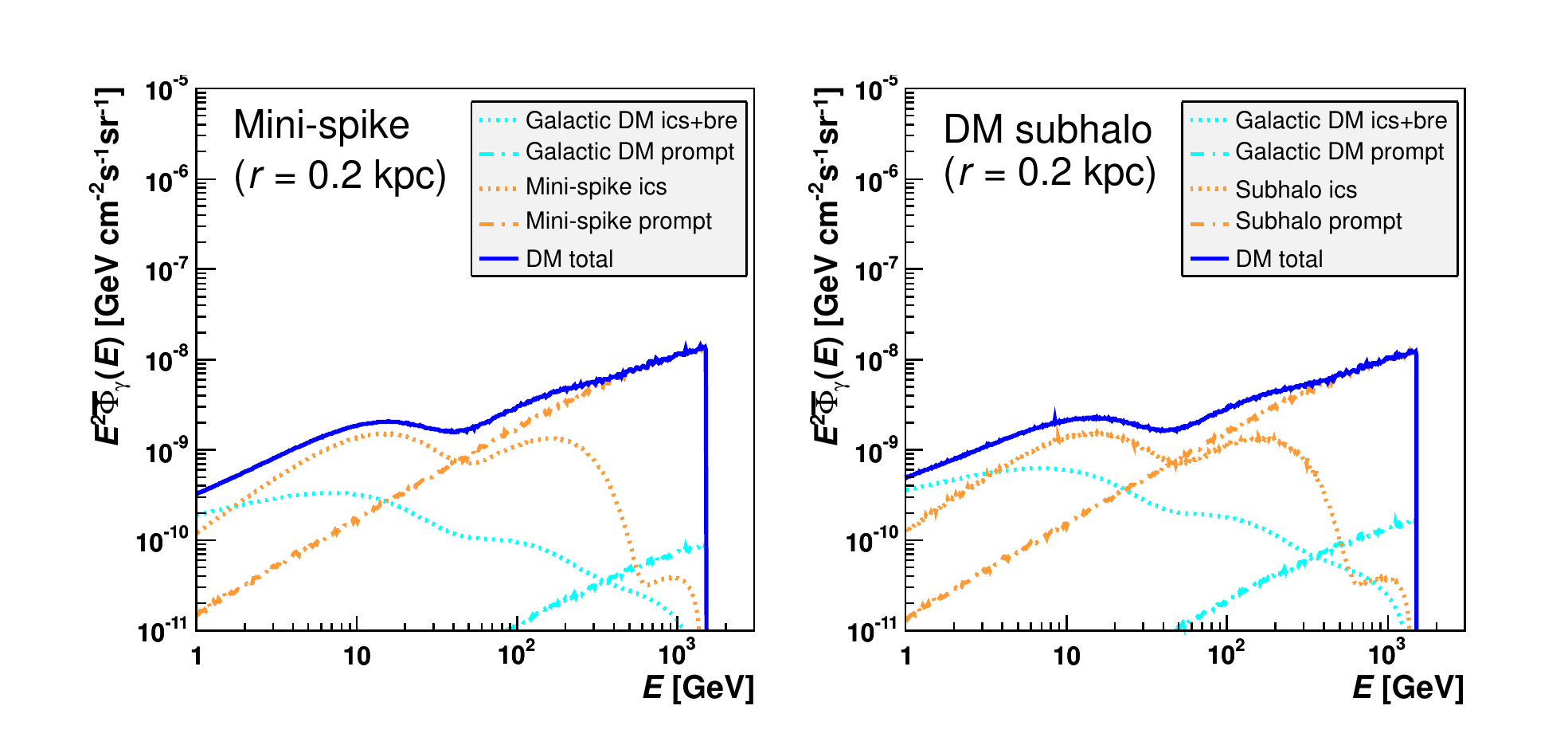}
\caption{
Left panels)
Top: 
spatial extension of the $\gamma$-ray flux (above 1~GeV) of 
the  mini-spike with parameters corresponding to the case of $r=0.2$~kpc in \tab{tab:pointDM}.
The contributions from prompt and ICS photons are shown explicitly, 
together with that from the Galactic halo DM.
Middle:
averaged energy spectrum of $\gamma$-rays over  
the ROI of a circular region with angular radius $30^{\circ}$ centered
at GC $(\ell=0^{\circ}, b=0^{\circ})$ 
for the same parameters.
Bottom:
the same as middle panels but for the ROI centered at  ($\ell=0^{\circ}$ and $b=90^{\circ}$).
Right panels)
The same as left, but for the DM subhalos with 
parameters corresponding to the case of $r=0.2$~kpc in \tab{tab:extendDM}.
}
\label{fig:GAMMA_DM_energy}
\end{center}
\end{figure}

In the top-left panel of \fig{fig:GAMMA_DM_energy}, 
we show the spatial extension of the $\gamma$-ray flux 
(with energy above 1~GeV) of the mini-spike
with the parameters corresponding to 
the best-fit values  in the case of $r=0.2$~kpc  in \tab{tab:pointDM}
and the direction of the location  coincides with the GC.
The corresponding DM annihilation cross section is 
$\langle \sigma v\rangle=1.36\times 10^{-26}~\text{cm}^{3}\text{s}^{-1}$ 
according to  Eq.~(3) in the letter. 
In the figure, the contributions of prompt and ICS photons are shown explicitly.
For mini-spikes, the spatial extension is $\ll 1^{\circ}$ within $68\%$ containment. 
Thus they can be treated as point-like sources.
For the three cases considered in \tab{tab:pointDM}, 
the total fluxes integrated in the energy range $1-100$~GeV
are in the range $(0.47-1.01)\times 10^{-10}~\text{cm}^{-2}\text{s}^{-1}$,
which are within the current sensitivity of Fermi-LAT in point-source searches.
For the point-source searches, the energy spectrum is dominated by the FSR photons.
Thus the expected spectrum should have a power index close to $\sim 1$. 
In the Fermi-LAT 3FGL catalog of unassociated point sources
(sources that have not been associated with emission observed at other wavelengths),
we find 6 candidate 
sources with low power indexes (index $\alpha$ can reach 1.0 withint $2\sigma$  error ), 
which are listed in the \tab{tab:3FGL}.
It is also possible that the mini-spike is located in the direction of the low Galactic
latitudes where the Fermi-LAT sensitivity  is much lower. 
\begin{table}[htbp]
	\centering
		\begin{tabular}{l|cc|cccc|c} 
		\hline\hline
        Source Name & $\ell$(deg) & $b$(deg) & $\Phi\ (10^{-10}\text{cm}^{-2}\text{s}^{-1})$ & $\alpha$ & Significance ($\sigma$) \\
     \hline
     J0603.3+2042 & 189.124 & -0.690422 & $6.180$ & $1.50\pm0.50$ & 4.37156 \\
     J1250.2-0233 & 302.344 & 60.3066 & $0.926\pm0.521$ & $1.10\pm0.30$ & 5.12018 \\
     J2209.8-0450 & 55.6854 & -45.5583 & $1.784\pm0.974$ & $1.27\pm0.32$ & 6.46061 \\
     J1705.5-4128c & 345.052 & -0.281031 & $48.379$ & $2.77 \pm1.06$ & 5.12339 \\
     J2142.6-2029 & 31.1422 & -46.5567 & $1.245\pm0.683$ & $1.52\pm0.33$ & 4.05409 \\
     J2300.0+4053 & 101.243 & -17.2428 & $1.719\pm0.746$ & $1.51\pm0.26$ & 5.2434 \\
		\hline\hline
	 \end{tabular}%
	 \caption{
	 Parameters of selected point sources in the Fermi-LAT 3FGL  ``unassociated'' point-source catalog [35], 
	 which have power-law type of spectrum with low power indices (the power
	 index $\alpha$ can reach one with $2\sigma$ error) and are possibly due to  FSR of
	 DM particles in mini-spikes. The total flux is integrated from 
	 1 to 100 GeV. %
	}
	\label{tab:3FGL}%
\end{table}%

In the top-right panel of \fig{fig:GAMMA_DM_energy}, 
we show the spatial extension of the DM subhalo
with the parameters corresponding to 
the best-fit values  in the case of $r=0.2$~kpc of \tab{tab:extendDM},
and the direction of the location  coincides with the GC.
It can be seen that the $\gamma$-rays are significantly extended to 
a few tens of degrees, as they are dominated by the ICS photons.
In \fig{fig:GAMMA_DM_energy}, for a comparison, 
we also show the prompt and ICS photons from 
the Galactic halo DM with the same DM annihilation cross section,
which is calculated by using  GALPROP with the same parameter set
as that use in obtaining the \fig{fig:DM-Galactic-halo}.
Unlike the case of electrons, the contribution to  diffuse $\gamma$-ray
from the Galactic halo DM is very significant.

The $\gamma$-ray energy spectra of mini-spikes and DM subhalos 
are shown in the left and right panels of \fig{fig:GAMMA_DM_energy}, 
respectively. 
The spectra are averaged over two regions of interest (ROI)
which represent the low and high Galactic latitude limits,
one is a circular ROI with an angular radius $30^{\circ}$ centered
at the GC ($\ell=0^{\circ}$ and $b=0^{\circ}$), 
the other one has the same shape but centered at 
high latitude   ($\ell=0^{\circ}$ and $b=90^{\circ}$).
The components of prompt and ICS photons are explicitly shown.
The contributions from the Galactic halo DM are included in the same way.
At low latitudes  close to the Galactic disk
the contributions of ICS photons are significant,
which is expected as the ISRF has large intensity.
After summing up all the components of the contributions, 
the total spectrum scales with energy approximately as $E^{-2}$
with a sharp cutoff at the DM particle mass $m_{\chi}$.
On the other hand, at high latitudes, the FSR photons are dominant at
high energies, and the spectrum scales approximately as $E^{-1}$.

\begin{figure}[!htbp]
\begin{center}
\includegraphics[width=0.9\textwidth]{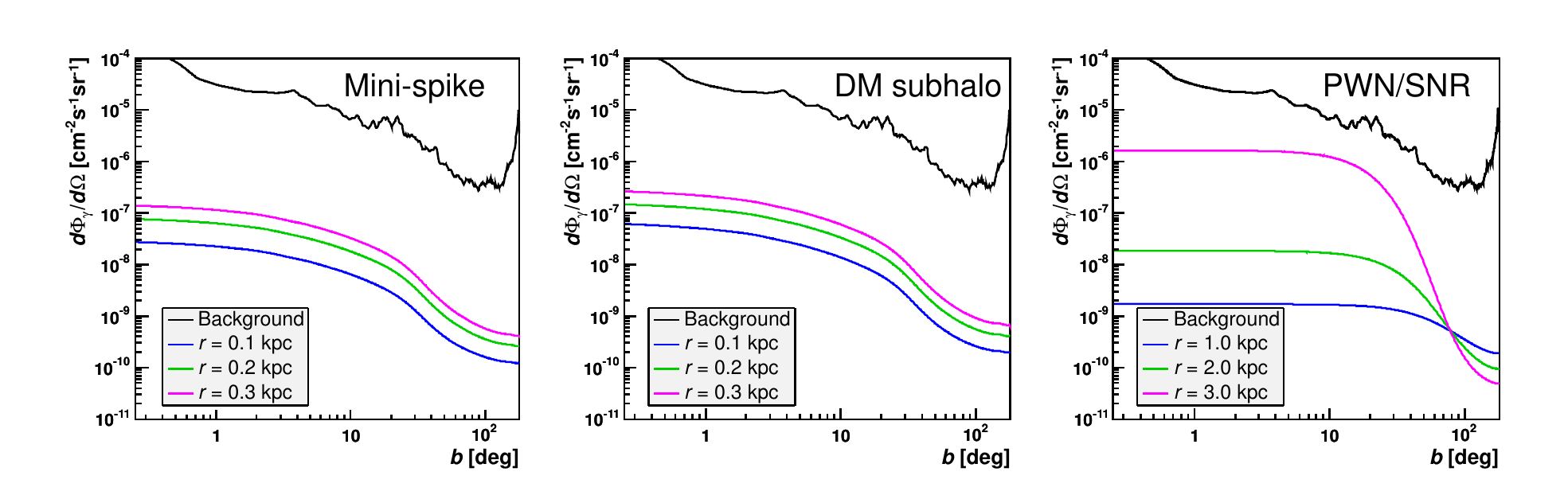}
\\
\includegraphics[width=0.9\textwidth]{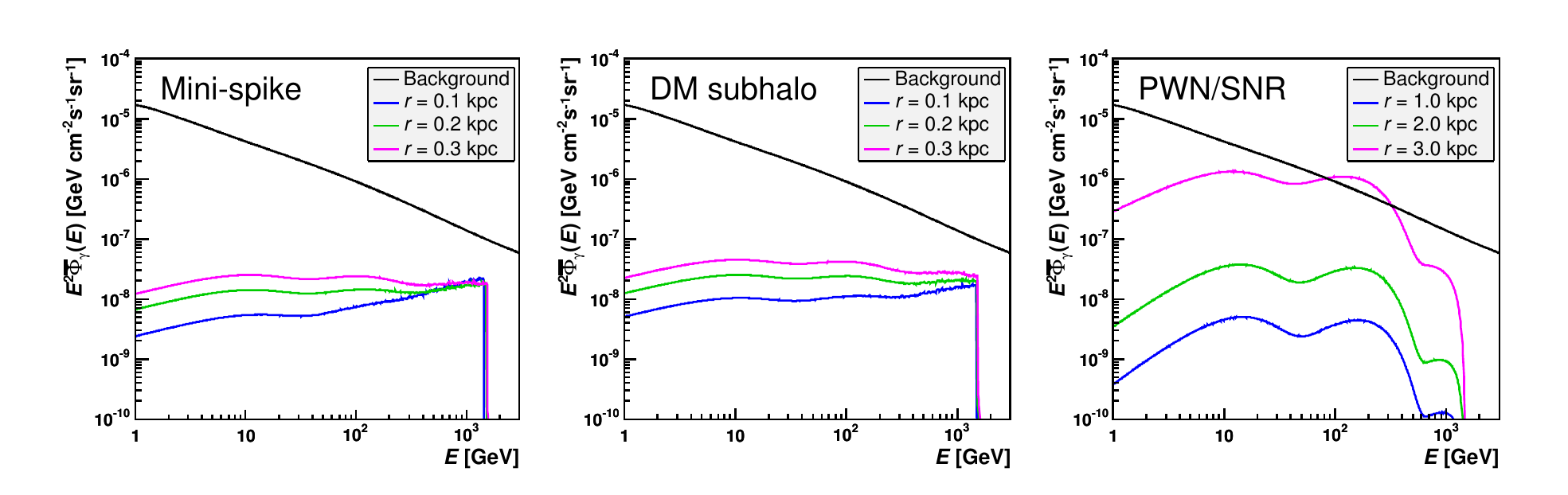}
\\
\includegraphics[width=0.9\textwidth]{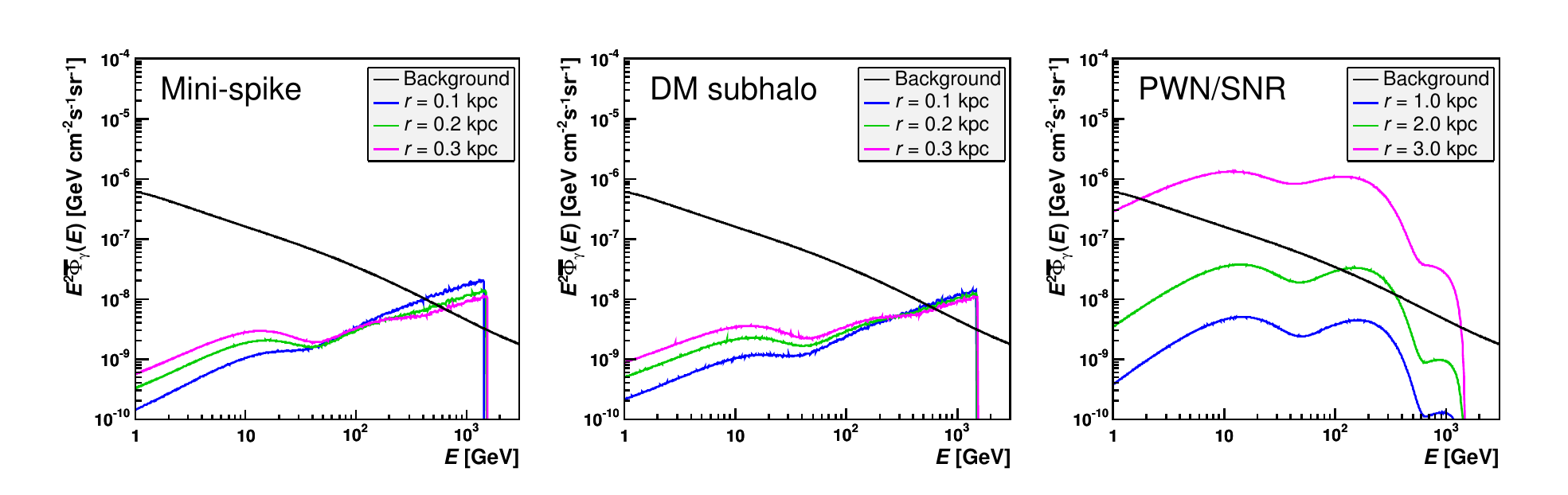}
\caption{
Predicted $\gamma$-ray spatial extension and energy spectra 
for all the three parameter sets in the three type of sources considered in 
Fig.~2 
in the letter with all the components summed up together.
Left panels) for mini-spikes with parameters in \tab{tab:pointDM}
Center panels) for DM subhalos with parameters in \tab{tab:extendDM}
Right panels) for burst-like sources with parameters in \tab{tab:SNR}.
The ROI are the same as that in \fig{fig:GAMMA_DM_energy}.
The corresponding Galactic diffuse $\gamma$-ray backgrounds are also shown.
}
\label{fig:GAMMA_cases_30}
\end{center}
\end{figure}

The spatial extension and energy spectra of $\gamma$-rays  
in all the cases of  the three type of sources 
listed in Tabs.~\ref{tab:pointDM}, \ref{tab:extendDM} and \ref{tab:SNR} are
summarized  in  \fig{fig:GAMMA_cases_30},
together with the calculated Galactic diffuse $\gamma$-ray backgrounds 
in the same ROIs.
The figure shows that at high latitude $b=90^{\circ}$, 
the predicted flux of burst-like source can exceed the diffuse background.
Since the background is in an overall agreement with the Fermi-LAT data,
this scenario is disfavoured. 
We thus conclude that the sources are more likely to be located 
at relatively low Galactic latitudes where the background is large.

